\documentstyle[12pt,epsfig]{article}
\def\lsim{\raise0.3ex\hbox{\,$<$\kern-0.75em\raise-1.1ex\hbox{$\sim$}\,}}
\def\gsim{\raise0.3ex\hbox{\,$>$\kern-0.75em\raise-1.1ex\hbox{$\sim$}\,}}

\begin{document}

\begin{titlepage}
\begin{flushright}
6 May, 2002\\
HIP-2002-23/TH\\
hep-ph/0205048
\end{flushright}
\begin{centering}
\vfill

{\bf A PERTURBATIVE QCD ANALYSIS OF CHARGED-PARTICLE DISTRIBUTIONS 
IN HADRONIC AND NUCLEAR COLLISIONS
}

\vspace{0.5cm}
 K.J. Eskola\footnote{kari.eskola@phys.jyu.fi}
and  H. Honkanen\footnote{heli.honkanen@phys.jyu.fi}\\

\vspace{1cm}
{\em Department of Physics, University of Jyv\"askyl\"a,\\
P.O.Box 35, FIN-40351 Jyv\"askyl\"a, Finland\\}
\vspace{0.3cm}
{\em Helsinki Institute of Physics,\\
P.O.Box 64, FIN-00014 University of Helsinki, Finland\\}
\end{centering}

\vspace{1cm} \centerline{\bf Abstract} 

We compute the distributions of charged particles at large transverse
momenta in $p\bar p(p)$, $pA$ and $AA$ collisions in the framework of
perturbative QCD, by using collinear factorization and the modern PDFs
and fragmentation functions.  At the highest cms-energies the shape of
the spectra measured in $p\bar p(p)$ collisions at large $q_T$ can be
well explained. The difference between the data and the lowest-order
computation is quantified in terms of a constant $K$-factor for each
energy.  The $K$-factor is found to systematically decrease with
growing $\sqrt s$.  Also a lower limit for the partonic transverse
momentum, $p_0$, is extracted for each $\sqrt s$ based on the
comparison with the measurements. A systematic increase of $p_0$ as a
function of $\sqrt s$ is found. Nuclear effects in the
charged-particle spectra in $pA$ and $AA$ collisions at RHIC and LHC
are studied in the framework of collinear factorization by applying
the EKS98 nuclear corrections to the parton distributions. The nuclear
effects are shown to mostly enhance the computed spectra. A comparison
with the recent PHENIX data from central and peripheral Au+Au
collisions at RHIC is done.

\vfill
\end{titlepage}

\section{Introduction}

Perturbative QCD (pQCD) and collinear factorization has been shown to be
successful in explaining the production of observable jets in
$p\bar p$ collisions \cite{ELLIS,ELLIS2}. The elements in such calculations
are the parton distribution functions (PDF) which contain
non-perturbative input at some initial scale $Q_0^2$, 
perturbatively computed cross sections for parton-parton scatterings,
and the infrared-safe measurement functions which determine the cross
section to be computed.  For the same energies it is, however,
surprisingly difficult to predict inclusive one-particle distributions
of charged hadrons in hadron-hadron collisions from perturbative QCD,
even at large cms-energies $\sqrt s\gsim 100$~GeV and $p_T\gg
\Lambda_{\rm QCD}$, where the perturbation theory should work very
well.

In the framework of collinear factorization, the inclusive
charged-particle transverse momentum spectra can be computed by convoluting the
inclusive production of a parton $k$ further with the corresponding
fragmentation function of $k$ into a hadron $h$,
$D_{k\rightarrow h}$. Schematically
\begin{equation}
d\sigma^{pp\rightarrow h+X} = \sum_{ijk} f_i(x_1,Q^2)
f_j(x_2,Q^2)\otimes \hat\sigma^{ij\rightarrow k+x}
(x_1,x_2,Q^2,\alpha_{\rm s}(\mu^2))
\otimes D_{k\rightarrow h}(z,\mu_{F}^2),
\end{equation}
where the partonic cross sections $\hat\sigma^{ij\rightarrow k+x}$ are
expressed as a power expansion in $\alpha_{\rm s}(\mu)$. The
factorization scale is denoted by $Q$, the renormalization scale
by $\mu$, and the fragmentation scale by $\mu_{F}$. State-of-the-art 
calculations where next-to-leading order (NLO) partonic cross
sections and NLO PDF are incorporated with NLO fragmentation functions
\cite{BKK,KKP} extracted from $e^+e^-$ collisions can be found in
\cite{BORZUMATI,AURENCHE}. While in these calculations the NLO results
are well compatible with the measured cross sections
\cite{UA1,MIMI,CDF} at large transverse momenta, the shape of the data
is generally not well reproduced at $q_T\lsim5$ GeV.  This can be
thought to be as a consequence of the fact that the fragmentation functions
of gluons are not sufficiently constrained by the $e^+e^-$ data, or
that the effects of intrinsic $k_T$ both in the fragmentation
functions and in the PDF are neglected. Especially at lower
cms-energies the $k_T$-effects are expected to become important, and
phenomenological approaches studying this problem starting from $pp$
collisions at $\sqrt s\sim 20 $ GeV have been developed e.g. in
\cite{XNW,LEVAI} based on the lowest-order (LO) pQCD approach.

In this paper, we will not make an effort to perform a NLO computation
of the charged-particle transverse momentum spectra, or to invoke a
phenomenological model for the intrinsic $k_T$-effects. Instead, we compute 
the inclusive production
of charged particles in LO but strictly remaining within the
collinearly factorized approach, supplementing the computation only
with a constant $K$-factor but no other phenomenological factors.  We
show that the shape of the computed transverse momentum distributions
agrees reasonable (surprisingly?) well with the measured ones in most
cases.  Emphasizing the highest values of transverse momenta measured
at each cms-energy, we determine the $K$-factors ($K\equiv
d\sigma^h_{\rm exp}/d\sigma_{\rm LO}^h$) based on a $\chi^2$ fit to
the data.  A systematic decrease of $K$ with growing cms-energy is found.

Particle production in hadronic and nuclear collisions is often
modelled in terms of minijet production above some transverse momentum
cut-off $p_0$, supplemented with a nonperturbative component for
$p_T\le p_0$ \cite{XNW91,SJOSTRAND,EKL}.  The lower limit of partonic
transverse momentum needs to be phenomenologically determined from the
measurements. This is done e.g. using the total and inelastic cross sections 
in $p\bar p$ and $pp$ scatterings in context with eikonal models
\cite{XNW91,newHIJING}. The charged-particle spectra measured at
midrapidities, however, offer another independent way of obtaining
constraints for the cut-off scale. Our second goal in
this paper is to try to find lower limits for $p_0$ based on the
comparison of the computed and measured $p_T$ distributions of charged
hadrons. A systematic increase of $p_0$ as a function of $\sqrt s$
is found.

Our third goal is to study the magnitude of the nuclear
effects in high-$p_T$ hadron production in $pA$ and $AA$ collisions at
RHIC and LHC energies by strictly keeping the framework of collinear
factorization and DGLAP evolution. For doing this we apply the EKS98
nuclear modifications \cite{EKS98} to the PDF, and investigate the relative
difference between $AA$ collisions and the corresponding $pp$ collisions.  
Both RHIC and LHC
energies are studied.  We show that the pQCD spectra in central Au+Au
collisions at RHIC at $q_T\lsim$ 10 GeV are enhanced by the nuclear effects 
at most by 15\%.

The fourth and final goal of the present paper is to compare the
computed pQCD spectra of charged hadrons to the very exciting recent
data from RHIC measured by PHENIX \cite{PHENIX}. To obtain the absolute
normalization for the spectra, we need to extract the $K$-factor for
$\sqrt s=130$ GeV based on the obtained behaviour $K(\sqrt s)$.  
Two centrality
classes are studied. The more peripheral one is found to be consistent
with $pp$ collisions, while in the 0...10\% central sample, the
computed transverse momentum spectra systematically lie above the data 
- especially at the highest transverse momenta, where the emphasis of our
approach is.

The paper is organized as follows. In Sec. 2 we present the formulae
for the computed pQCD cross sections in detail. Sec. 3 explains how the 
$K$-factors are extracted and summarizes the $p\bar p(p)$ results. 
In Sec. 4 the 
study is extended to nuclear collisions. Sec. 5 contains the
comparison with the PHENIX data. Discussion is given in Sec. 6.

\section{The formulae}

The formulae for computing inclusive charged-particle
production in LO pQCD are given below. 
In LO, the partons are produced back-to-back in the 
transverse plane according to the differential cross section
\begin{equation}
\frac{d\sigma}{dp_T^2dy_1dy_2}^{\hspace{-0.6cm}AB\rightarrow kl+X} =
\sum_{ij}x_1 f_{i/A}(x_1,Q^2) x_2f_{j/B}(x_2,Q^2) \frac{d\hat\sigma}{d\hat
t}^{ij\rightarrow kl}\hspace{-0.6cm}(\hat s,\hat t,\hat u)
\label{2parton}
\end{equation}
where $A$ and $B$ denote the colliding hadrons or nuclei.  
The rapidities of the final state partons $k$
and $l$ are labelled by $y_1$ and $y_2$, and transverse momentum of
each parton by $p_T$. The fractional momenta of the colliding partons
$i$ and $j$ are $x_{1,2}=\frac{p_T}{\sqrt s}({\rm e}^{\pm y_1}+{\rm e}^{\pm
y_2})$, i.e. the incoming partons are collinear with the beams. The
parton distributions are obtained from the CERN-PDFLIB library
\cite{PDFLIB}, and the factorization scale is $Q\sim p_T$.  The
Mandelstam variables for the subprocesses are denoted by $\hat s,\hat t$
and $\hat u$. In LO, the partonic cross sections are
$d\hat\sigma/d\hat t\sim \alpha_{\rm s}^2(\mu^2)$. The 1-loop expression for
the strong coupling constant is used here, so
\begin{equation}
\alpha_{\rm s}(\mu^2)= 
\frac{12\pi}{(33-2N_f)\log(\mu^2/\Lambda_{\rm QCD}^{(N_f)\,\,\,2})},
\label{alphas}
\end{equation}
where $\mu\sim p_T$ is the renormalization scale.  In the present
study all partons are considered massless and the number of active
flavours, $N_f=5,4,3$ below the corresponding threshold scales
$Q_6=m_t=174$ GeV, $Q_5=m_b=4.5$ GeV and $Q_4=m_c=1.5$ GeV.  For
$N_f=4$, we will use the $\Lambda_{\rm QCD}^{(4)}$ as given by PDFLIB
\cite{PDFLIB} for the chosen set of parton distributions. The change
of $\Lambda_{\rm QCD}^{(N_f)}$ is computed by requiring matching of 
$\alpha_{\rm s}$ at the threshold scales.  The cross
sections for the eight different partonic subprocesses can be found
e.g. in \cite{SARCEVIC}.

Inclusive cross section for production of a parton of a flavour $f$ and a
rapidity $y_f$ is obtained by integrating over one of the rapidities
in Eq.~(\ref{2parton})
and keeping track of the parton flavours \cite{EK},
\begin{eqnarray}
\frac{d\sigma}{dp_T^2 dy_f}^{\hspace{-0.4cm}AB\rightarrow f+X}\hspace{-0.6cm}
&=& \int dy_1 dy_2 \sum_{\langle kl\rangle}
\frac{d\sigma}{dp_T^2dy_1dy_2}^{\hspace{-0.6cm}AB\rightarrow kl+X} 
\hspace{-0.5cm}
\left[ \delta_{kf} \delta(y_f-y_1) + \delta_{lf}\delta(y_f-y_2) \right]
\frac{1}{1+\delta_{kl}}
\\ \nonumber
&=&\int dy_2 \sum_{\langle ij\rangle \langle kl\rangle} 
\frac{1}{1+\delta_{kl}}\frac{1}{1+\delta_{ij}} \times
\\\nonumber
&&\times\Bigg\{  x_1f_{i/A}(x_1,Q^2)\,x_2f_{j/B}(x_2,Q^2) \left[
\frac{d\hat\sigma}{d\hat t}^{ij\rightarrow kl}
\hspace{-0.7cm}(\hat s,\hat t,\hat u)
\delta_{fk} + \frac{d\hat\sigma}{d\hat t}^{ij\rightarrow kl}
\hspace{-0.7cm}(\hat s,\hat u,\hat t)
\delta_{fl} \right] \\
&&+ x_1f_{j/A}(x_1,Q^2)\,x_2f_{i/B}(x_2,Q^2) \left[
\frac{d\hat\sigma}{d\hat t}^{ij\rightarrow kl}
\hspace{-0.7cm}(\hat s,\hat u,\hat t) \delta_{fk} + 
\frac{d\hat\sigma}{d\hat t}^{ij\rightarrow kl}
\hspace{-0.7cm}(\hat s, \hat t,\hat u)\delta_{fl} \right] 
\Bigg\}
\label{1parton}
\end{eqnarray}
where the summations run over the pairs, i.e.  $\langle
kl\rangle,\langle ij\rangle = gg,gq,g\bar q,qq,q\bar q,q\bar q, \bar
q\bar q$, ($q=u,d,s,\dots$)
without a mutual change.

For inclusive hadron production through fragmentation of
the parton $f$, let us define the transverse momentum of a hadron $h$ be $q_T$,
its transverse mass  $m_T$ and rapidity $y$. We define the fraction $z$
to be the ratio of the energy of the hadron and the energy of the
parton $f$ from which the hadron originates: $z=E_h/E_f$. The hadron
is considered to be produced collinearly with the parton. With these 
definitions the partonic variables become related to the hadronic ones 
through the relations
\begin{equation}
m_T\cosh y=zp_T\cosh y_f\quad\quad {\rm and}\quad\quad 
m_T\sinh y = q_T\sinh y_f.
\end{equation}
The inclusive cross section for hadron production can then be written
as
\begin{eqnarray}
\nonumber 
\frac{d\sigma}{dq_T^2dy}^{\hspace{-0.3cm}AB\rightarrow h+X}
=\frac{d\sigma}{dm_T^2dy}^{\hspace{-0.3cm}AB\rightarrow h+X}
\hspace{-0.3cm}&=& K(\sqrt s)\sum_f
\int_{p_0^2} dp_T^2 dy_f \frac{d\sigma^{AB\rightarrow f+X}}{dp_T^2 dy_f}
\int_0^1 dz D_{f\rightarrow h}(z,\mu_F^2)\\
&&\times\delta(m_T^2-M^2_T(q_T,y_f,z))\delta(y-Y(q_T,y_f,z))
\label{fragm}
\end{eqnarray}
where the transverse mass and rapidity of the hadron are
expressed in terms of the integration variables as 
\begin{equation}
M^2_T(p_T,y_f,z) = (zp_T)^2+m^2\tanh^2y_f, \quad\quad\quad 
Y(p_T,y_f,z) = {\rm arsinh}(\frac{q_T}{m_T}\sinh y_f).
\label{kinematics}
\end{equation}
For the fragmentation functions $D_{f\rightarrow h}$ we shall use
the set KKP \cite{KKP} which is the latest one. The fragmentation 
scale is $\mu_F\sim q_T$. 

One of the tasks below will be to determine the $K$-factor based  
on charged-hadron production measured in $p\bar p(p)$ collisions at 
collider energies in the range  $\sqrt s=63\dots1800$ GeV. 
The $K$-factor thus effectively accounts for the higher-order effects 
in the partonic cross sections, in the running coupling, in the parton 
distributions and fragmentation functions, and also for the possible 
intrinsic transverse momentum effects in the parton showers both in 
the initial and in the final state. We shall analyze the $\sqrt s$
dependence of $K$, in particular.

The parameter $p_0$ in Eq.~(\ref{fragm}) is the smallest transverse
momentum of parton scatterings allowed. For models with semi-hard
parton production, this parameter plays a key role. We shall also
discuss the lower limit for $p_0$ for each $\sqrt s$.

For completeness, we write down the cross section above in an explicit
form, which can be numerically evaluated in a straightforward manner.
Performing the integrations over $p_T^2$ and $y_f$ in Eq.~(\ref{fragm})
gives
\begin{eqnarray}
\frac{d\sigma}{dq_T^2dy}^{\hspace{-0.3cm}AB\rightarrow h+X}
&=& K(\sqrt s)\cdot J(m_T,y)
\sum_f\int\frac{dz}{z^2}\, 
D_{f\rightarrow h}(z,\mu_F^2) 
\frac{d\sigma^{AB\rightarrow f+X}}{dp_T^2 dy_f}\bigg|_{p_T^2,y_f} \\\nonumber
&=& K(\sqrt s)\cdot J(m_T,y) \int\frac{dz}{z^2}
\int dy_2 \sum_{\langle ij\rangle \langle kl\rangle} 
\frac{1}{1+\delta_{kl}}\frac{1}{1+\delta_{ij}} \times
\\\nonumber
&&\hspace{-3.5cm}\Bigg\{  x_1f_{i/A}(x_1,Q^2)\,x_2f_{j/B}(x_2,Q^2) \left[
\frac{d\hat\sigma}{d\hat t}^{ij\rightarrow kl}
\hspace{-0.7cm}(\hat t,\hat u)
D_{k\rightarrow h}(z,\mu_F^2) + \frac{d\hat\sigma}{d\hat t}^{ij\rightarrow kl}
\hspace{-0.7cm}(\hat u,\hat t)
D_{l\rightarrow h}(z,\mu_F^2) \right] \\
&&\hspace{-3.5cm}+ x_1f_{j/A}(x_1,Q^2)\,x_2f_{i/B}(x_2,Q^2) \left[
\frac{d\hat\sigma}{d\hat t}^{ij\rightarrow kl}
\hspace{-0.7cm}(\hat u,\hat t) D_{k\rightarrow h}(z,\mu_F^2) + 
\frac{d\hat\sigma}{d\hat t}^{ij\rightarrow kl}
\hspace{-0.7cm}(\hat t,\hat u)D_{l\rightarrow h}(z,\mu_F^2) \right] 
\Bigg\}
\label{dsigma/dqT2}
\end{eqnarray}
where the partonic cross section is evaluated at
\begin{equation}
p_T=\frac{q_T}{z}J(m_T,y),\quad y_f={\rm arsinh}(\frac {m_T}{q_T}\sinh y)
\end{equation}
and where 
\begin{equation}
J(m_T,y) = (1-\frac{m^2}{m_T^2\cosh^2 y})^{-1/2}.
\label{jaakoppi}
\end{equation}
The integration region for $y_2$, $-\log(\sqrt s/p_T-{\rm e}^{-y_f})\le 
y_2\le \log(\sqrt s/p_T-{\rm e}^{y_f})$, is over the whole phase space, 
whereas that
for $z$,
\begin{equation}
\frac{2m_T}{\sqrt s}\cosh y \le z
\le {\rm min} \left[ 1, \frac{q_T}{p_0}J(m_T,y) \right],
\end{equation}
is affected by the requirement $p_T\ge p_0$. In this way, 
we shall also be able to  consider the region $q_T<p_0$. 

The experiments most often give the cross sections averaged over 
a pseudorapidity interval $\Delta\eta$, defined as
\begin{equation}
\frac{d\sigma^h_{AB}}{dq_T^2d\eta}\bigg|_{\eta\in\Delta\eta}
\equiv \frac{1}{\Delta\eta}\int_{\Delta\eta} d\eta 
\frac{d\sigma^h_{AB}}{dq_T^2d\eta} 
= \frac{1}{\Delta\eta}\int_{\Delta \eta} d\eta J(m_T,y)^{-1}
 \frac{d\sigma^h_{AB}}{dq_T^2dy}, 
\label{pseudo}
\end{equation}
where $y={\rm arsinh}(\frac{q_T}{m_T}\sinh\eta)$ and 
$J(m_T,y)^{-1}=\partial y/\partial\eta$, where $J(m_T,y)$ is the same
factor as in  Eq.~(\ref{jaakoppi}).

\section{The analysis and results for $p\bar p(p)$}

The approach of collinear factorization and independent fragmentation 
can be expected to work best at the highest values of $q_T$ and at the 
highest cms-energies. We determine
$K$ for each $\sqrt s$ based on the data in the large-$q_T$
region by minimizing the $\chi^2$:
\begin{equation}
\chi^2(N) = \sum_{i=1}^N \left(
\frac{ K\sigma_i^{\rm th}-\sigma_i^{\rm exp}}{\Delta_i^{\rm exp} }\right)^2
\end{equation}
where $\sigma_i^{\rm exp}$ is the measured cross 
section at a transverse momentum $q_{Ti}$, with a statistical error  
$\Delta_i^{\rm exp} $. The cross sections computed at 
$q_{Ti}$ without the $K$-factor, are denoted by $\sigma_i^{\rm th}$. 
The $N$ data points included in the fit are counted starting from the 
highest $q_T$ measured for each $\sqrt s$. As will be seen,
the $K$-factor usually depends on $N$.

\begin{figure}[f]
\vspace{-2.5cm}
\centerline{\hspace{-2.cm} 
\epsfysize=17cm\epsffile{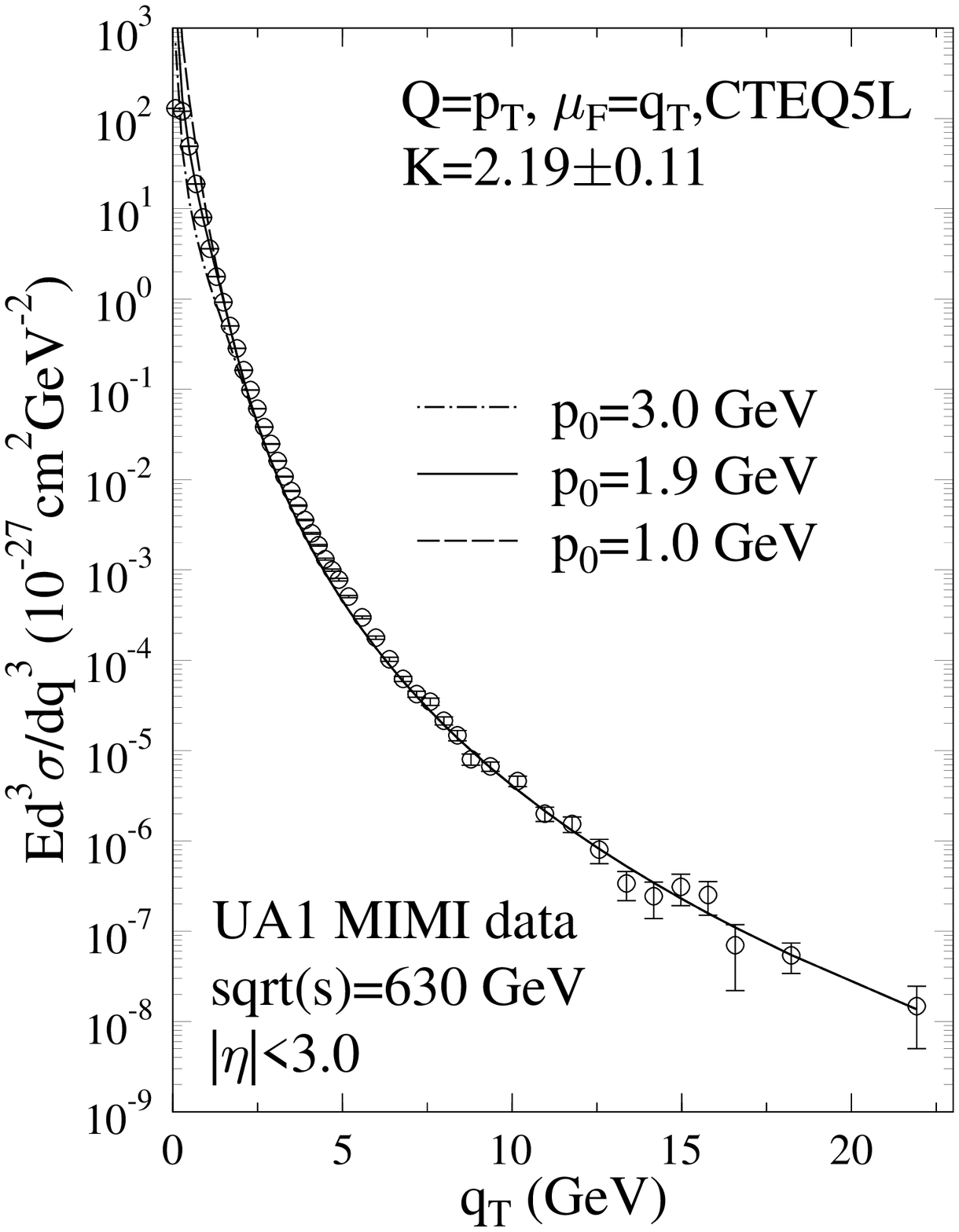} \hspace{-1.0cm}
\epsfysize=13.5 cm\epsffile{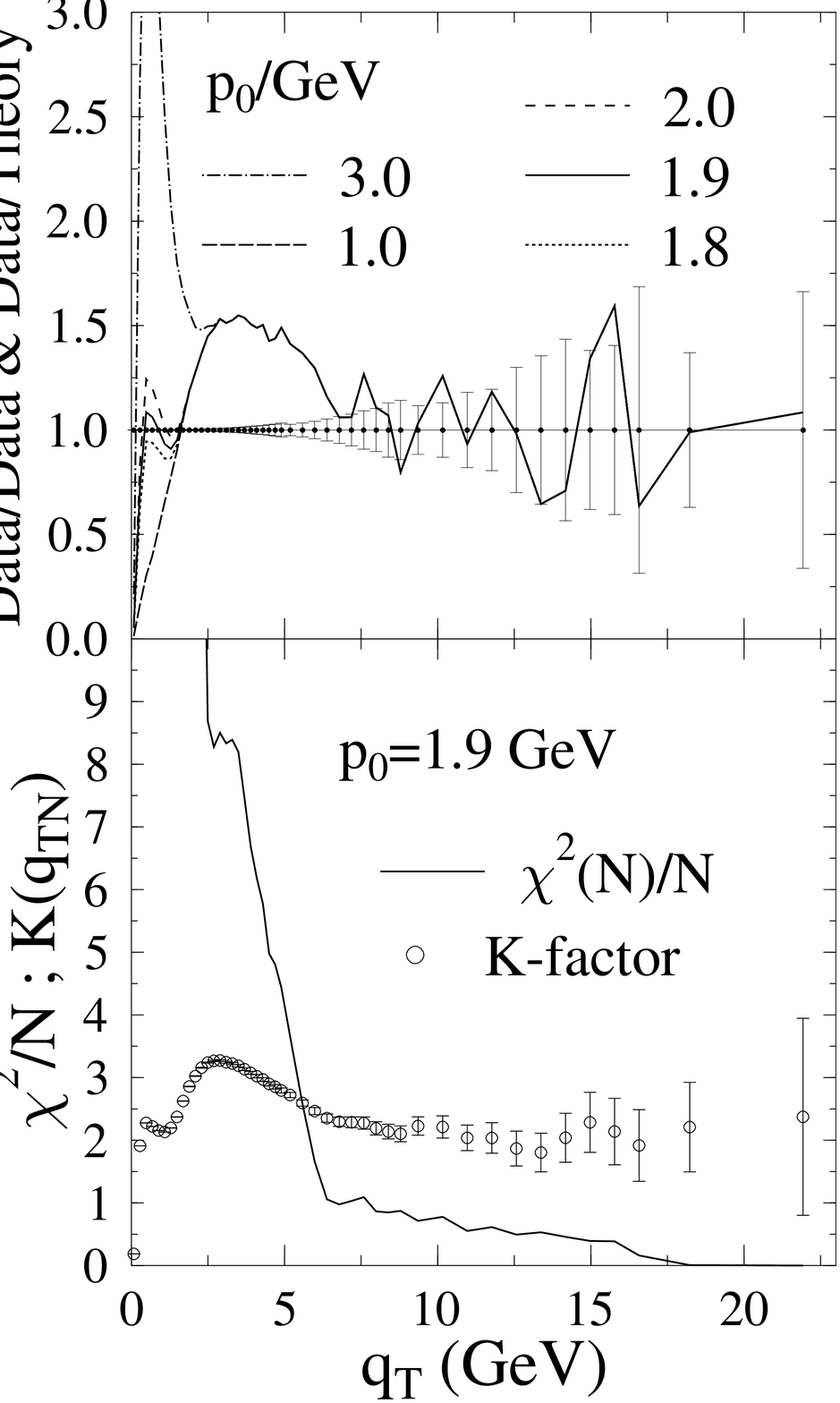} }
\vspace{-2cm}
\caption[a]{\protect \small {\bf Left}: Inclusive cross section for
charged-hadron production ($ h\equiv h^+ + h^- $) in $p\bar p$
collisions at $\sqrt{s}= 630$ GeV averaged over $\vert\eta\vert<3.0$.
The LO pQCD prediction with $K=2.19$ and scales $Q=p_T$, $\mu_F =q_T$
and $p_0=1.0, 1.9, 3.0$ GeV is shown by the curves. The data shown
with the statistical error bars are from UA1 MIMI \cite{MIMI}. {\bf Top
right}: The ratios data-to-data and data-to-theory as a function of
transverse momentum for various $p_0$. {\bf Bottom right}: The mimimized
$\chi^2(N)/N$ (solid curve) and the resulting $K$-factor (with error
bars) as a function of the smallest transverse momentum, $q_{TN}$,
included in the fit. The $K$-factor is read off from the point where
$\chi^2(N)/N=1$. The errors of the $K$-factor are computed from
Eq.~(\ref{k-error}).  }
\label{mitt630_1}
\end{figure}

Since in the region of large $q_T$ one is sensitive to the
(statistical) fluctuations in the experimental data, it would be
preferable to include as many data points in the fit as possible (most
preferably all points of course).  The previous comparisons
\cite{BORZUMATI, AURENCHE} of the pQCD cross sections with the data
indicate, however, that the pQCD results tend to fall somewhat below
the data in the region of $q_T\sim 3\dots5$ GeV. Also, the pQCD
cross sections clearly become too large at $q_T\lsim 1$ GeV, which
region is sensitive to the cut-off parameter $p_0$. The statistical
error bars of the data in the few-GeV region are clearly smaller than
those at large-$q_T$. Thus, if the few-GeV region is included in the
fit for the determination of the $K$-factor, a bad fit for the
large-$q_T$ region results.  This can be seen in Fig.~\ref{mitt630_1}
(bottom right), where we plot the $K$-factors obtained for the UA1
MIMI data at $\sqrt s=630$ GeV as a function of the smallest
transverse momentum $q_{TN}$ included in the fit.  In the same figure,
we also show $\chi^2/N$ as a quantitative measure of the goodness of
the fit.

To circumvent the problems mentioned above, and to make sure that the
$K$-factor gets determined from the region where the collinearly
factorized pQCD approach is supposed to work best, we determine the
maximum number of data points that can be included by requiring a good
fit, $\chi^2(N)/N\le 1$, at $q_T\ge q_{TN}$, and fix the $K$-factor 
based on these $N$ data points. We estimate the statistical significance 
of the $K$-factor at each $\sqrt s$ by computing an error \cite{NRECIPES}
\begin{equation}
\Delta_K(N) = \left[
\sum_{i=1}^N (\Delta_i^{\rm exp})^2
\left( \frac{\partial K}{\partial \sigma_i^{\rm exp}} 
\right)^2 \right]^{\frac{1}{2}}.
\label{k-error}
\end{equation}
This gives the error bars shown in Fig.~\ref{mitt630_1}, lower right.
For the UA1 MIMI data we find that $\chi^2(N)/N\leq 1.0$ 
at $q_T\geq 7.988$ GeV, with $N=15$, and  $K=2.19\pm 0.11$.

The $q_T$ spectrum of charged hadrons in $p\bar p$ at $\sqrt s=630$
GeV, computed from Eq.~(\ref{pseudo}) by applying the $K$-factor
determined above, is shown in Fig.~\ref{mitt630_1} (left panel)
together with the UA1 MIMI data for $p +\bar p\to h + X$ ($h\equiv
h^++ h^-$). The sum of charged hadrons always includes pions, kaons
and protons.  Both the data and the computation are averaged over
$-3.0<\eta<3.0$. Contrary to what perhaps could have been expected,
the agreement with the data are relatively good, taking into account
that we discuss particle production which extends over 10 orders of
magnitude. A more detailed comparison is done in the top right panel
of Fig.~\ref{mitt630_1}, where we plot the data divided by the
computed cross section, together with (relative) statistical error
bars of the data. From this figure we can see why large $\chi^2/N$
follows if the few-GeV region is included in the fit.

Fig.~\ref{mitt630_1} (left and top right) also illustrates how we can
estimate the lower limit for the parameter $p_0$: we require that the
computation must not overshoot the measured cross section in the
region $q_T\ge p_0$. For this particular data set, we notice that
e.g. for $p_0=3$ GeV the computed spectrum remains below the measured
one until the very last data point $q_T=0.079$ GeV.  The value $p_0=1$
GeV in turn would cause an overestimate of the cross section already
at $q_T\sim 2$ GeV. This search procedure leads to a lower limit 
$p_0=1.8\dots 2.0$ GeV for $\sqrt s=630$ GeV.

\begin{figure}[f]
\vspace{-2.5cm}
\centerline{\hspace{-2.cm} 
\epsfysize=17cm\epsffile{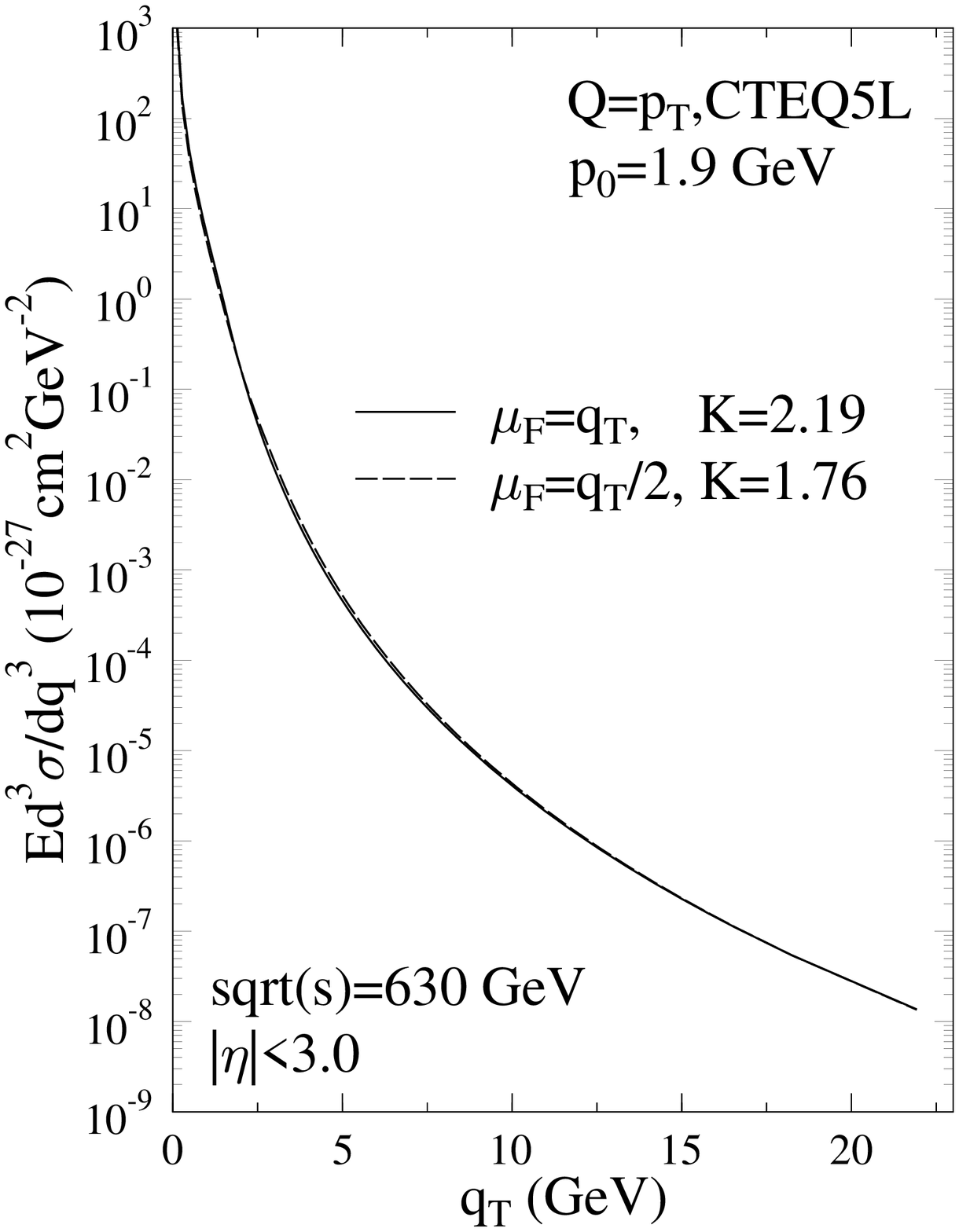} \hspace{-1.0cm}
\epsfysize=13.5cm\epsffile{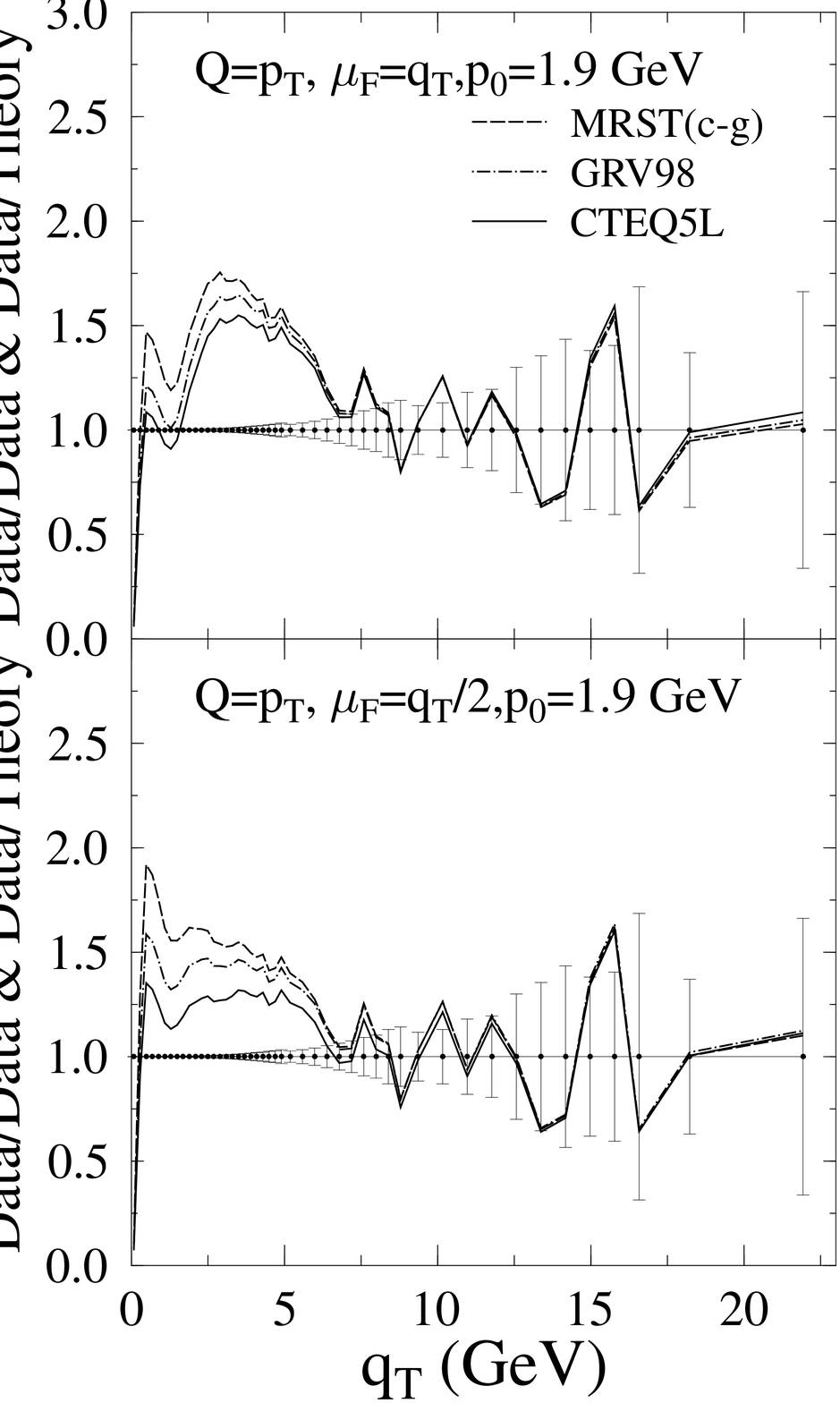} }
\vspace{-2cm}
\caption[a]{\protect \small {\bf Left}: An example of the dependence of the
inclusive pQCD cross sections (as in Fig.~\ref{mitt630_1}) on the
choice of the fragmentation scale $\mu_F$.  Based on the fit to the
large-$q_T$ region of the UA1 MIMI data in Fig.~\ref{mitt630_1}, we
obtain $K=2.19\pm 0.11$ for $\mu_F=q_T$ and $K=1.76\pm 0.05$ for
$\mu_F=q_T/2$.  The scale $p_0$ has been fixed to 1.9 GeV and CTEQ5 PDF
are used.  {\bf Right}: As in the top right panel of Fig.~\ref{mitt630_1} but
showing the dependence of the pQCD results on the PDF set.  Three
different sets are used, and the scales are fixed to $Q=p_T$ and
$p_0=1.9$ GeV in both panels. For the sets CTQE5L, GRV98 and
MRST(c-g), we obtain $K=2.19\pm0.11, 2.00\pm 0.10, 2.15\pm 0.10$ for
$\mu_F=q_T$ (upper panel) and $K=1.76\pm 0.05, 1.53\pm 0.07, 1.65\pm
0.08$ for $\mu_F=q_T/2$ (lower panel).}
\label{mitt630_1_scales}
\end{figure}

Sensitivity of the computed cross sections to the choice of the
fragmentation scale $\mu_F$ and to the different sets of parton
distributions is studied in Fig.~\ref{mitt630_1_scales} for the UA1
MIMI data.  The results are shown for the LO sets CTEQ5 \cite{CTEQ},
MRST(c-g) \cite{MRST}, GRV98 \cite{GRV98} with the scales $\mu_F=q_T$
and $\mu_F=q_T/2$.  Notice that in this figure we have already
included the $K$-factor which is determined as explained above. The
actual value of the $K$-factor is always correlated with these
choices, and for e.g. CTEQ5 and $\mu_F=q_T/2$, we get $K=1.76\pm
0.05$. We notice that qualitatively the behaviour of the computed
results relative to the data is the same as for
$\mu_F=q_T$. Quantitatively, the overall agreement between the
computed and measured cross sections could be somewhat improved by
optimizing the choice of the fragmentation scale and to some extent
also by a choice of the parton distributions (modern sets are to be
used, of course).  For $\mu_F=q_T$ the difference between the best and
worst fitting curve is $\sim 20\%$ and for $\mu_F=q_T/2$ $\sim
30-40\%$.  Although not studied here in more detail, there is also
some freedom for choosing the factorization and renormalization
scales, which could improve the quality of the overall fit. We do not,
however, pursue this study into the direction of optimizing the scale
choices based on fits to the data but choose to use $Q=\mu=p_T$,
$\mu_F = q_T$ and the CTEQ5 parton distributions with
$\lambda^{(4)}_{QCD}= 192$ MeV in what follows.

\begin{table}
\center
\begin{tabular}{|c|c|c|c|c|c|c|c|}
\hline
$\sqrt s$/GeV	&collab.	& ref.		& syst. err.	&$K\pm\Delta K$ 	&$p_0$/GeV	&$q_{TN}$/GeV	&N\\
\hline
63	&AFS		&\cite{AFS}	&20\%	     	&$6.19\pm 0.58$		&-		&4.65		&2\\
200	&UA1		&\cite{UA1}	&15\%		&$3.42\pm 0.40$		&1.6	        &3.45       	&8\\
500	&UA1		&\cite{UA1}	&15\%		&$1.94\pm 0.20$		&1.7		&3.35		&9\\
630	&UA1/MIMI	&\cite{MIMI}	&		&$2.19\pm 0.11$		&1.9		&7.988		&15\\
630	&CDF		&\cite{CDF}	&		&$1.78\pm 0.32$		&2.0	        &2.9      	&2\\
900	&UA1		&\cite{UA1}	&15\%		&$1.09\pm 0.10$		&1.6            &4.9  		&10\\
1800	&CDF		&\cite{CDF}	&		&$1.28\pm 0.18$		&2.2            &6.0  		&5\\     
\hline
\end{tabular}
\caption[1]{\protect\small The data sets used in the
analysis. Systematic errors are quoted when given. The $K$-factors
are obtained by requiring a good fit at the high-$q_T$ part of the
spectra (see the text), and the error $\Delta K$ is computed from
Eq.~(\ref{k-error}). Systematic error is not included in the estimates
for $\Delta K$. Notice that these $K$-factors are obtained with the
CTEQ5 \cite{CTEQ} set of parton distributions, and KKP \cite{KKP} set
of fragmentation functions and with the scale choices $\mu_F=q_T$,
$Q=p_T$.  The values of $q_{TN}$ and $N$ are also shown (see the
text).  }
\end{table}

The data sets we study are listed in Table~1. The systematic errors
given by the experiments are quoted. Table~1 also summarizes our
results for the $K$-factors along with their error estimates
(systematic errors are not included), and the corresponding $q_{TN}$
and $N$, as well as the obtained values of the lower limits of the cut-off 
scale $p_0$.

\begin {figure} [f]
\begin{center}
\includegraphics[totalheight=14cm]{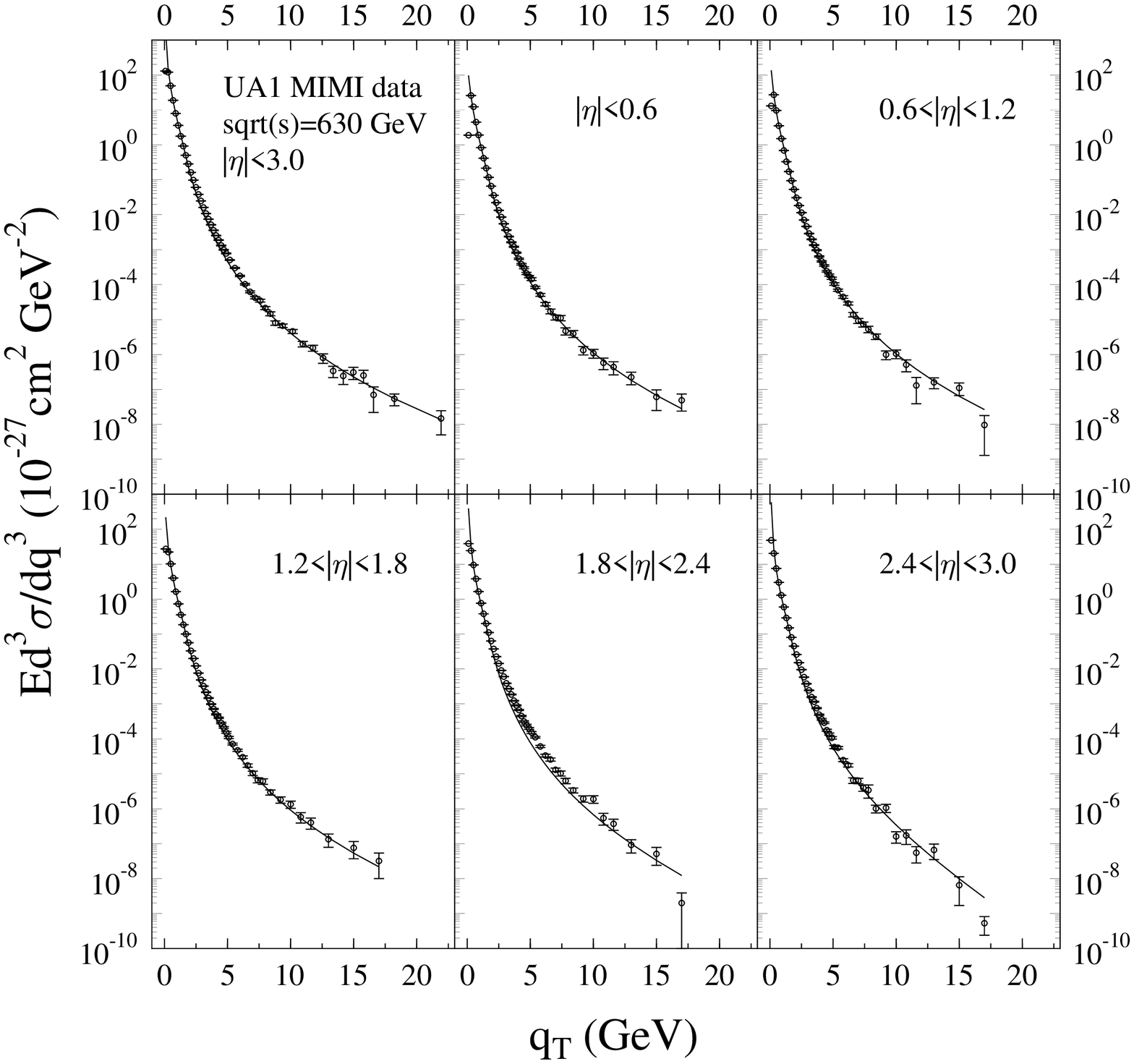}
\end{center}
\vspace*{-0cm}
\caption[a]{\protect \small The LO pQCD prediction for the charged-particle 
spectra in $p\bar p$ collisions at $\sqrt s=630$ GeV with $K=2.19$ and 
$p_0=1.9$ GeV. The upper left panel corresponds to Fig.~\ref{mitt630_1},
in the other panels the pseudorapidity interval $\vert\eta\vert <3.0$
has been divided into different subintervals. The data are from 
UA1 MIMI, Ref.~\cite{MIMI}.
}
\label{mitt630_all}
\end {figure}

Fig.~\ref{mitt630_all} presents the pQCD results with the UA1 MIMI
data at $\sqrt{s}=630$ GeV averaged over various pseudorapidity
intervals. We obtained $p_0=1.9$ GeV and $K=2.19\pm 0.11$ using the
data averaged over a range $\vert\eta\vert<3.0$. We use these values
to decompose the $q_T$ spectra into pseudorapidity intervals
$\vert\eta\vert<0.6$, $0.6<\vert\eta\vert<1.2$,
$1.2<\vert\eta\vert<1.8$, $1.8<\vert\eta\vert<2.4$,
$2.4<\vert\eta\vert<3.0$. As seen in the figure, these results also agree 
nicely with the
data except for $1.8<\vert\eta\vert<2.4$.

\begin{figure}[f]
\vspace{-2.0cm}
\centerline{\hspace{-2.cm} 
\epsfysize=17cm\epsffile{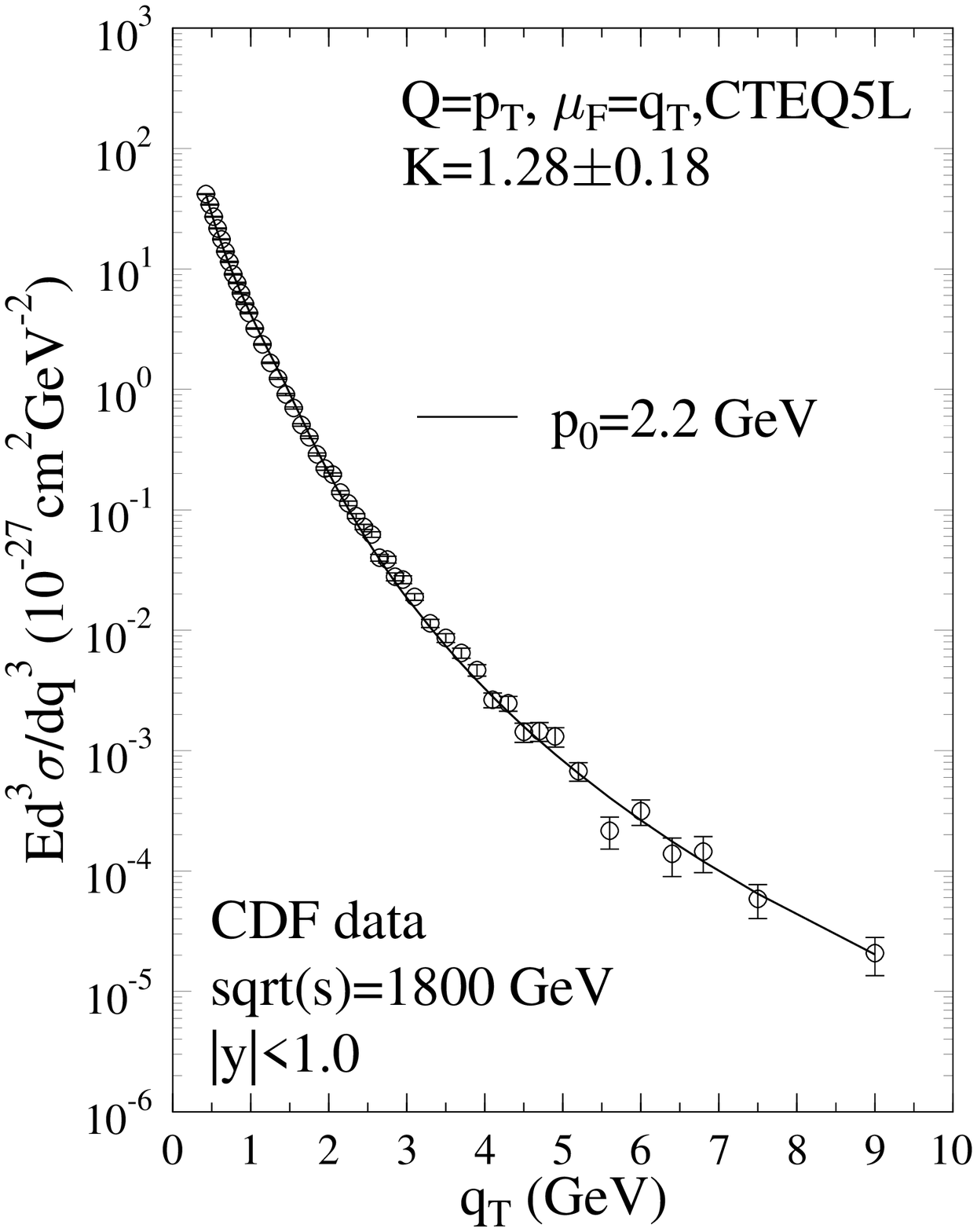} \hspace{-1.0cm}
\epsfysize=13.5cm\epsffile{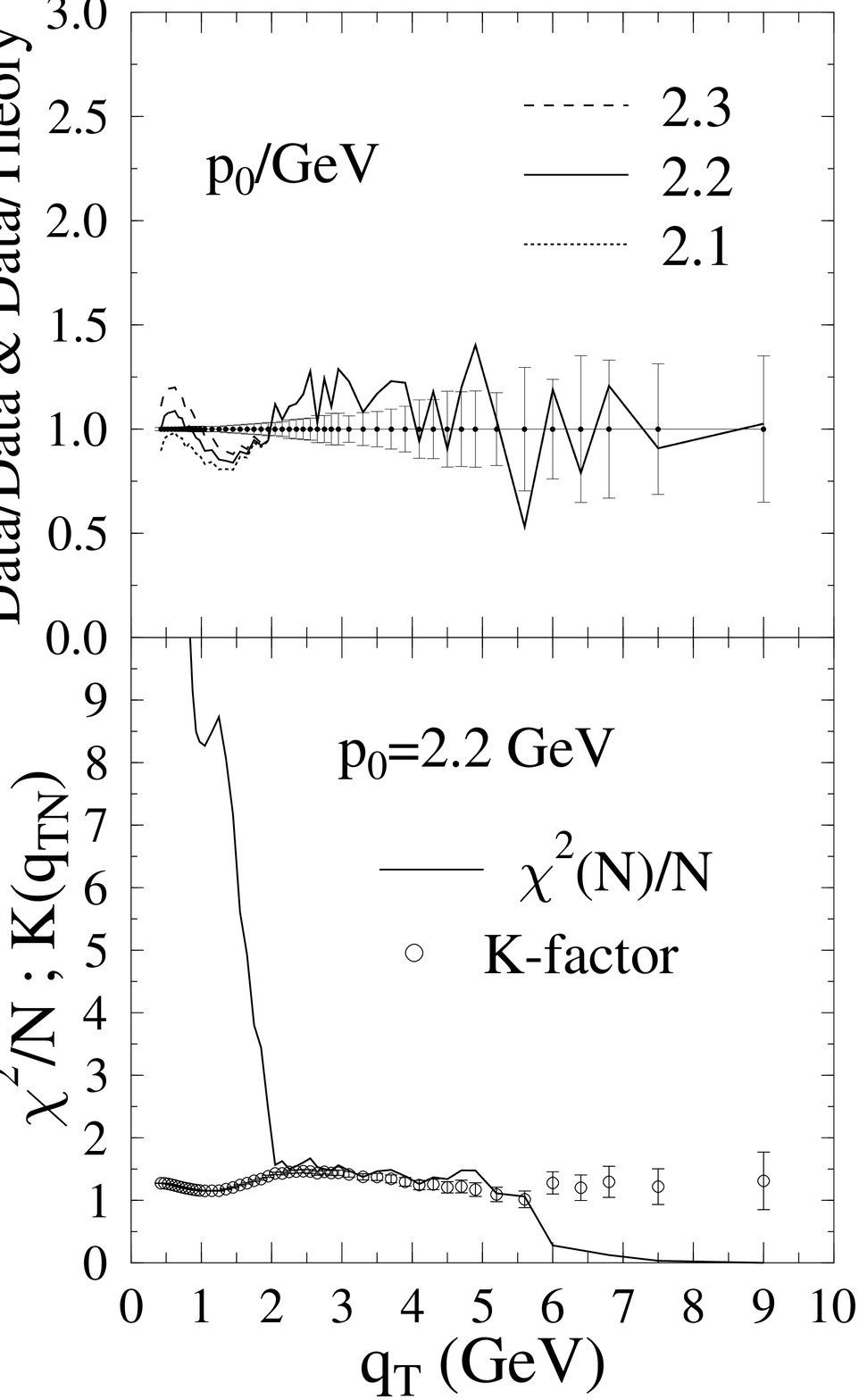} }
\vspace{-2cm}
\caption[a]{\protect \small As Fig.~\ref{mitt630_1} but for 
$h\equiv (h^++h^-)/2$, $\sqrt{s}=1800$ GeV and $\vert y\vert <1.0$.
The data are from CDF \cite{CDF}.
}
\label{mitt1800}
\end{figure}

\begin{figure}[f]
\vspace{-2.0cm}
\centerline{\hspace{-2.cm} 
\epsfysize=17cm\epsffile{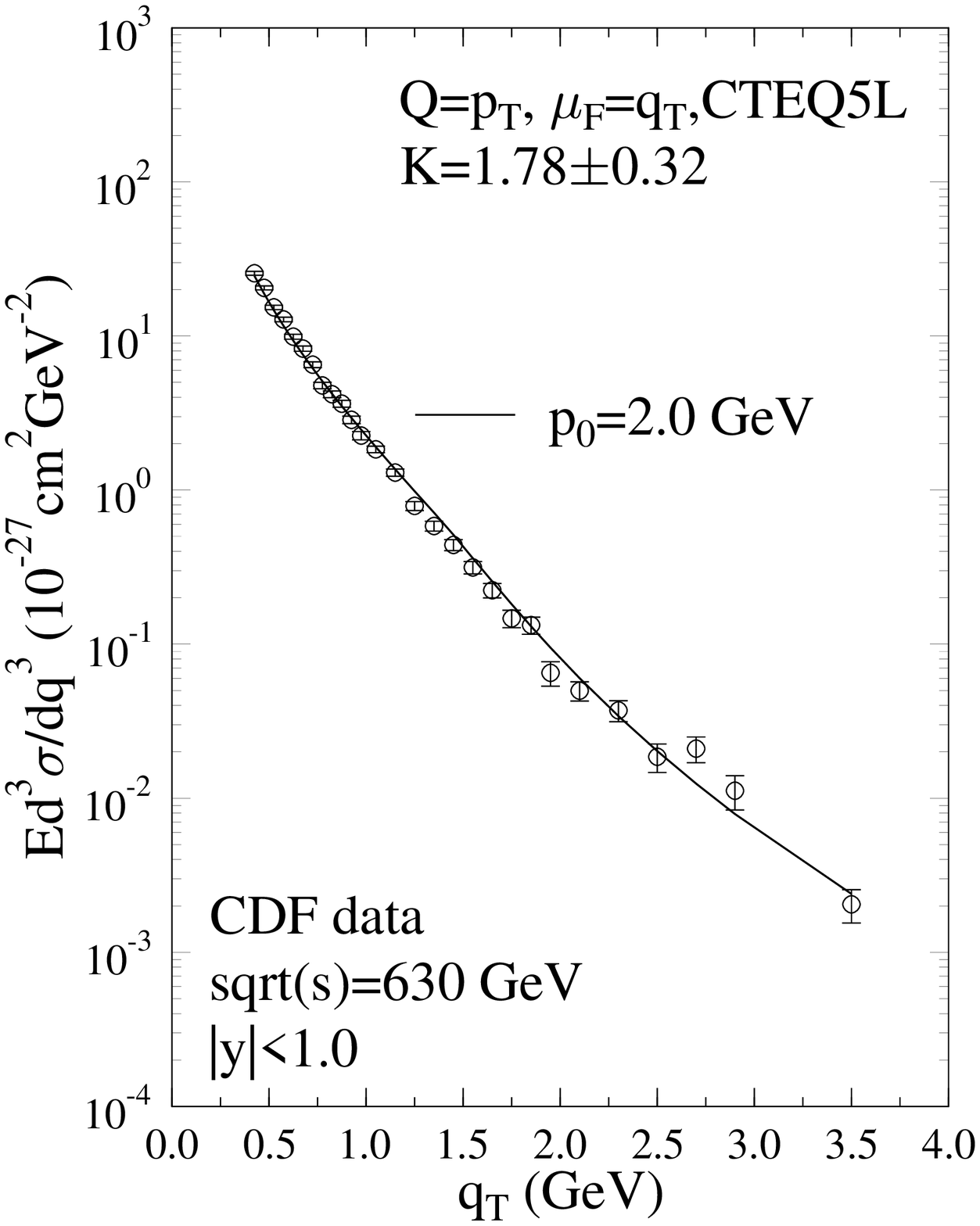} \hspace{-1.0cm}
\epsfysize=13.5cm\epsffile{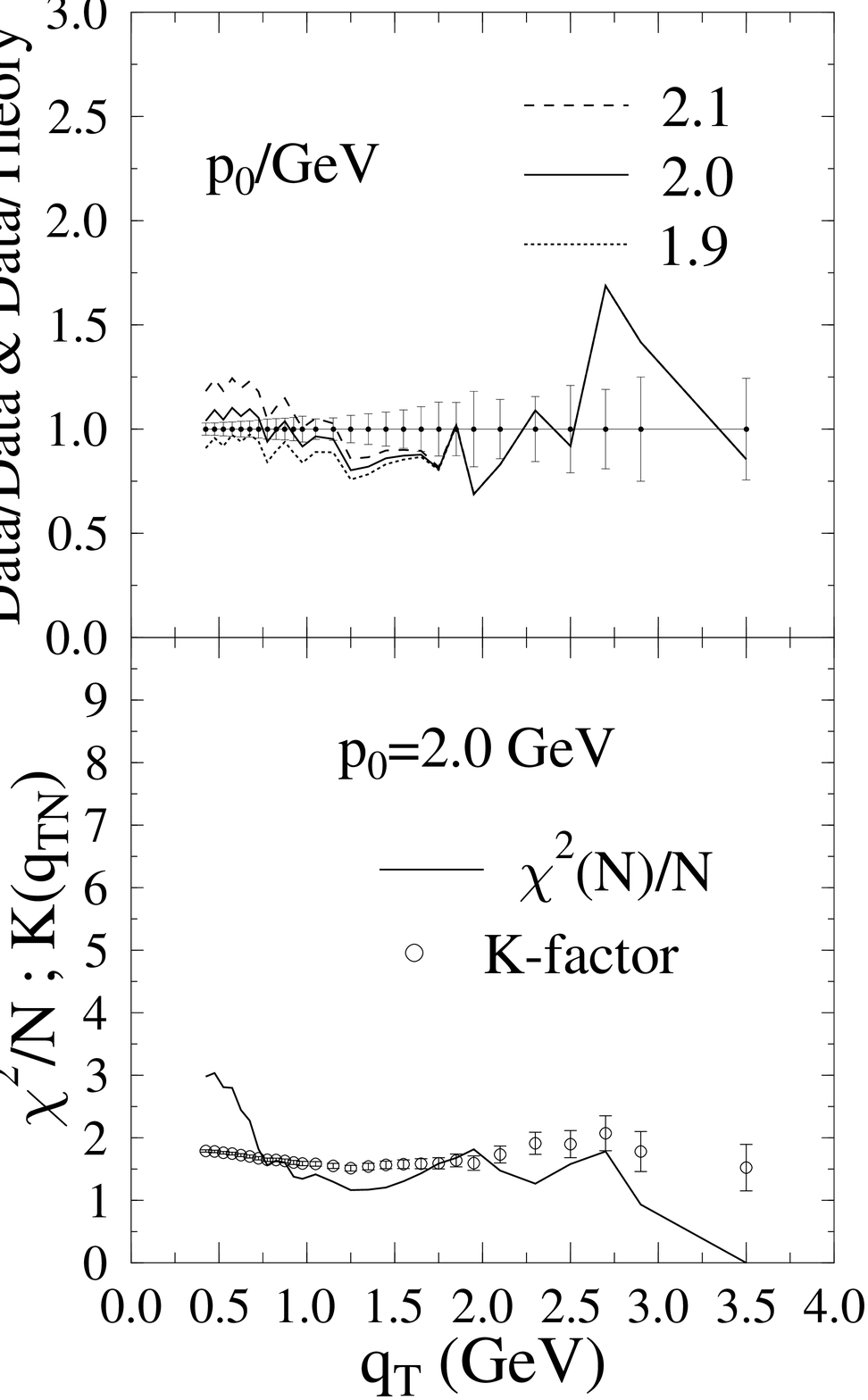} }
\vspace{-2cm}
\caption[a]{\protect \small  As Fig.~\ref{mitt1800} but for 
$\sqrt{s}=630$ GeV.  
}
\label{mitt630}
\end{figure}

Figs.~\ref{mitt1800} and \ref{mitt630} show the computed inclusive
charged-hadron cross sections for $p +\bar p\to h + X$ ($h\equiv h^++
h^-$) at $\sqrt{s}=$1800 and 630 GeV together with the CDF data
\cite{CDF}.  Averaging is done over a rapidity interval $\vert
y\vert<1.0$.  For $\sqrt{s}=1800$ GeV (Fig.~\ref{mitt1800}, bottom
right) the distribution of the obtained $K(q_{TN})$ is nearly flat
which indicates a good agreement with the data. Fig.~\ref{mitt1800},
top right, shows that this is indeed the case.  The computed absolute
spectrum is shown in the left panel of Fig.~\ref{mitt1800}. The
corresponding analysis for $\sqrt{s}=630$ GeV is shown in
Fig.~\ref{mitt630}. The agreement with the data seems to be again fairly
good although the $K$-factor is now fixed only by the last two
data points. Note that the data only cover a very limited $q_T$-range, 
$0.425<q_T<3.5$,  so in practice half of the computed spectrum 
is controlled by $K$ and the other half by $p_0$.

\begin{figure}[f]
\vspace{-2.0cm} \centerline{\hspace{-2.cm}
\epsfysize=17cm\epsffile{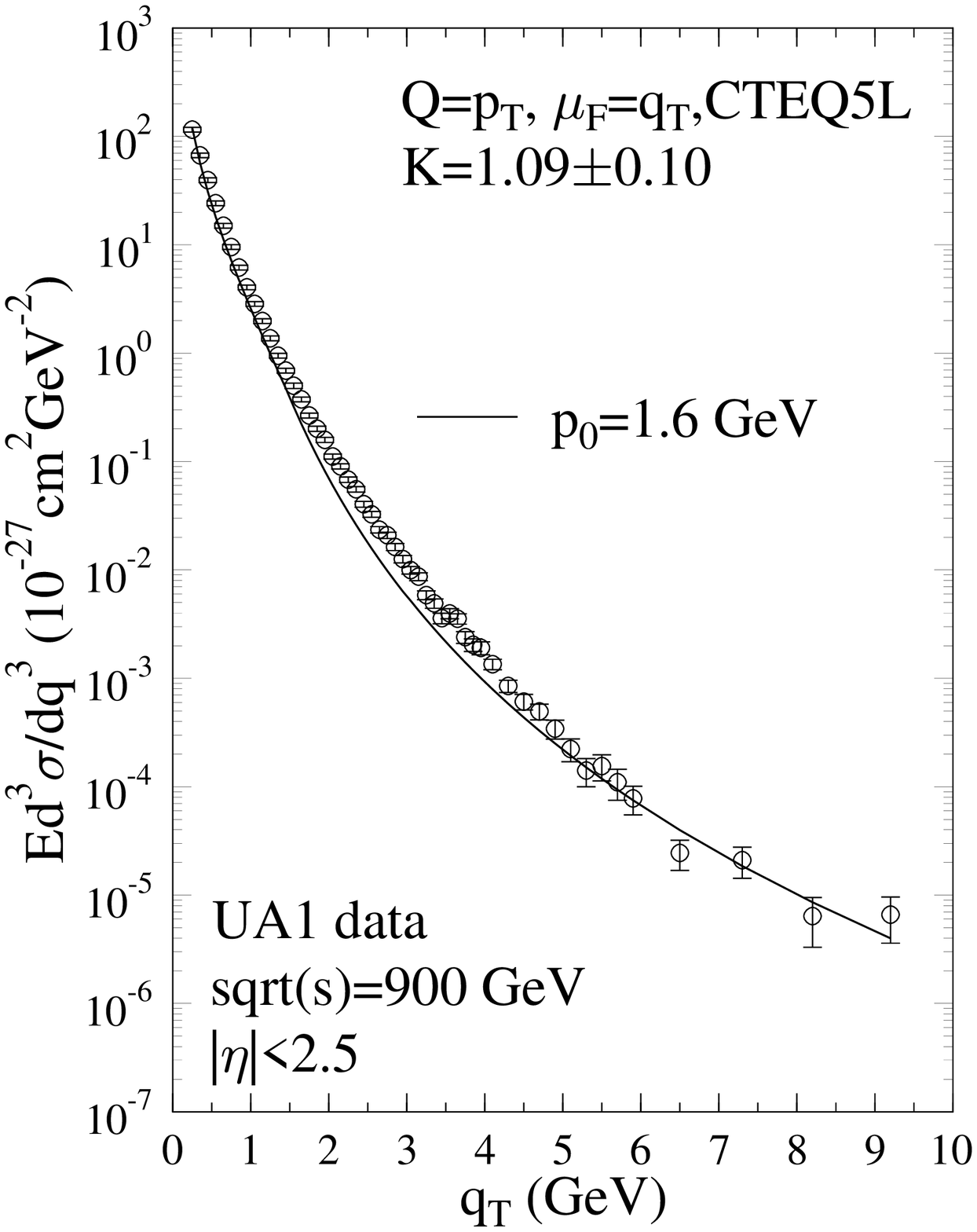} \hspace{-1.0cm}
\epsfysize=13.5cm\epsffile{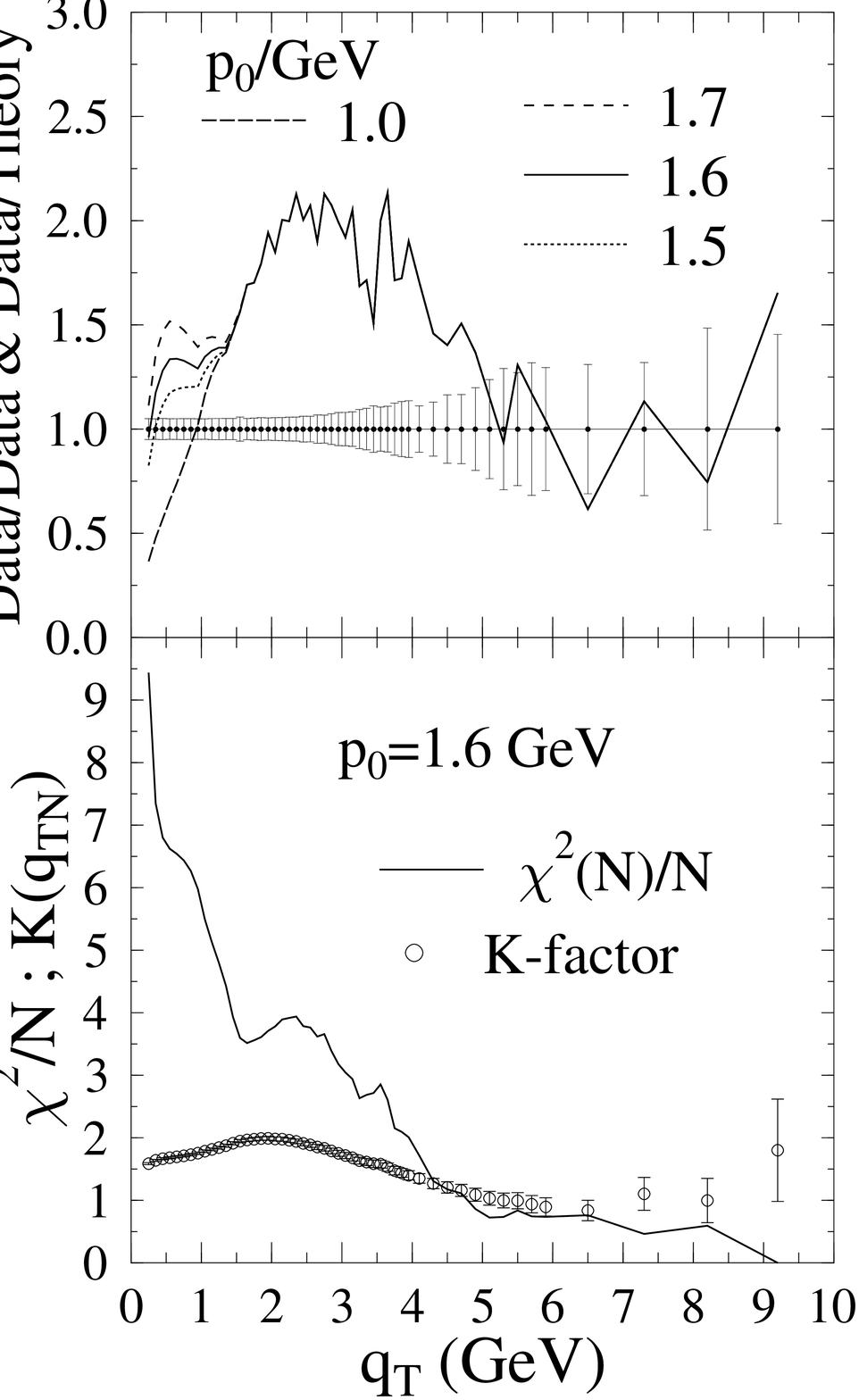} }
\vspace{-2cm}
\caption[a]{\protect \small 
As Fig.~\ref{mitt630_1} but for $h\equiv (h^+ +h^-)/2$,
$\sqrt{s}=900$ GeV and $\vert\eta\vert <2.5$.
The data are from UA1 \cite{UA1}.
}
\label{mitt900}
\end{figure}

\begin{figure}[f]
\vspace{-2.0cm} \centerline{\hspace{-2.cm}
\epsfysize=17cm\epsffile{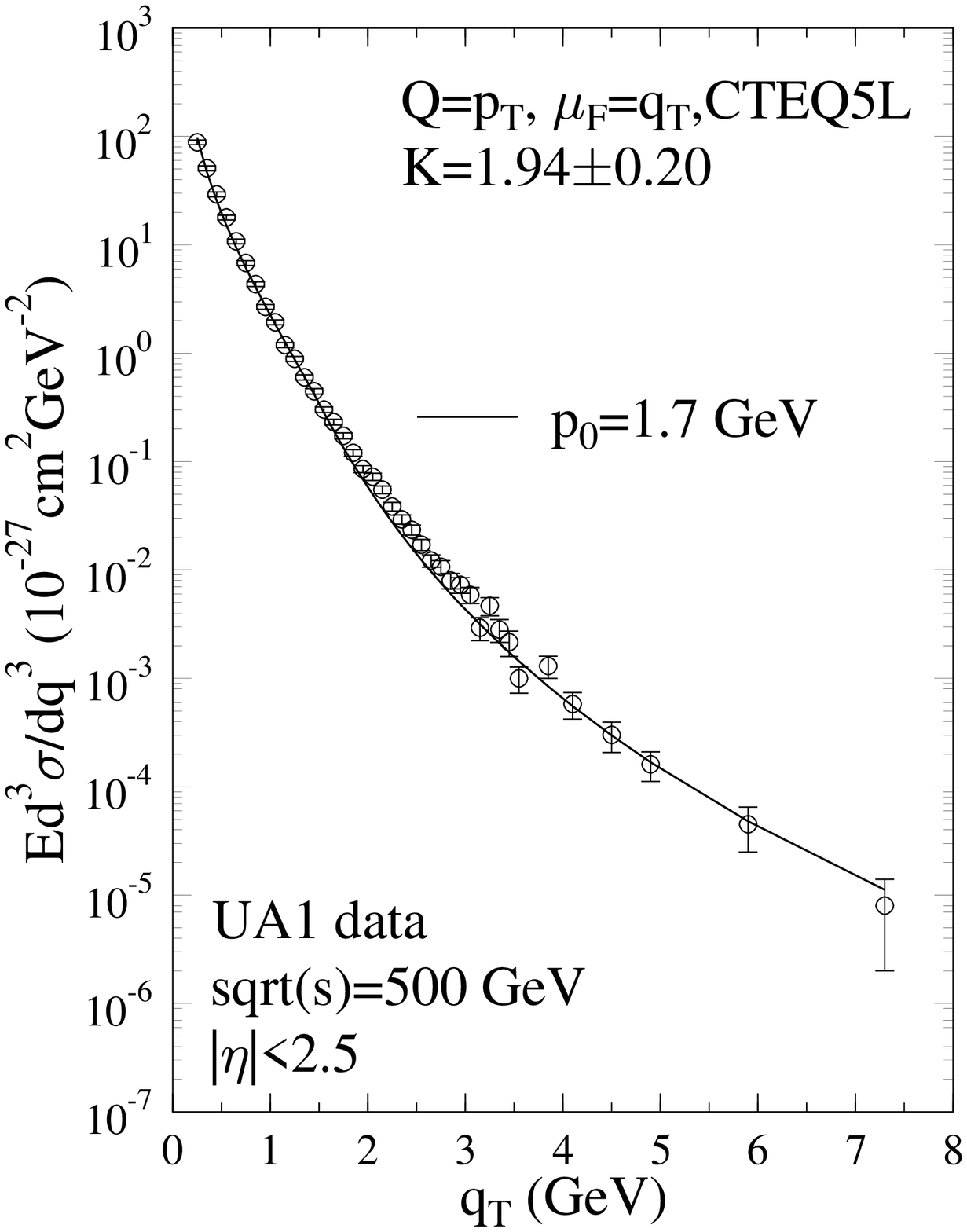} \hspace{-1.0cm}
\epsfysize=13.5cm\epsffile{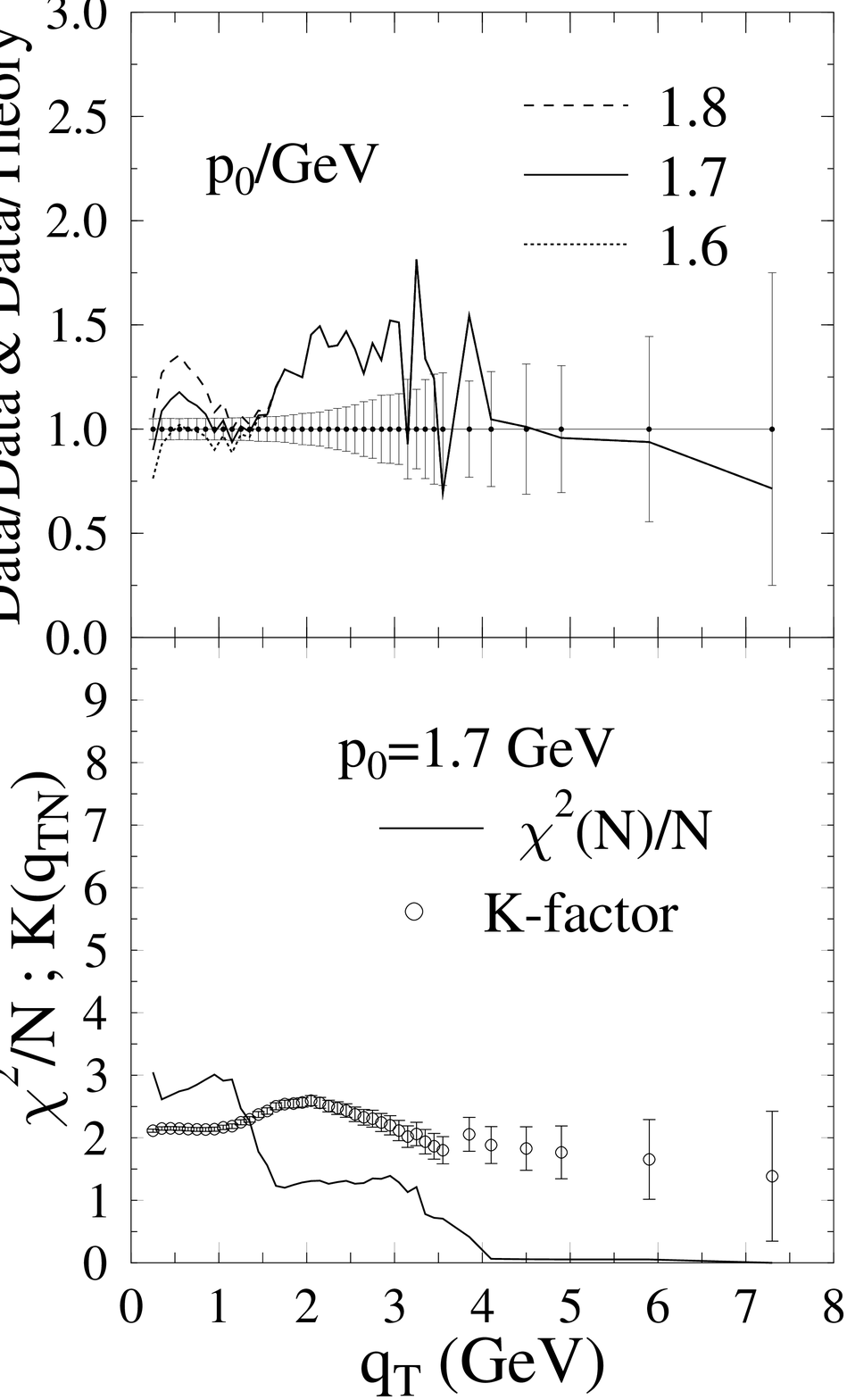} }
\vspace{-2cm}
\caption[a]{\protect \small As Fig.~\ref{mitt900} but for
$\sqrt{s}=500$ GeV.  
}
\label{mitt500}
\end{figure}

\begin{figure}[f]
\vspace{-2.0cm} \centerline{\hspace{-2.cm}
\epsfysize=17cm\epsffile{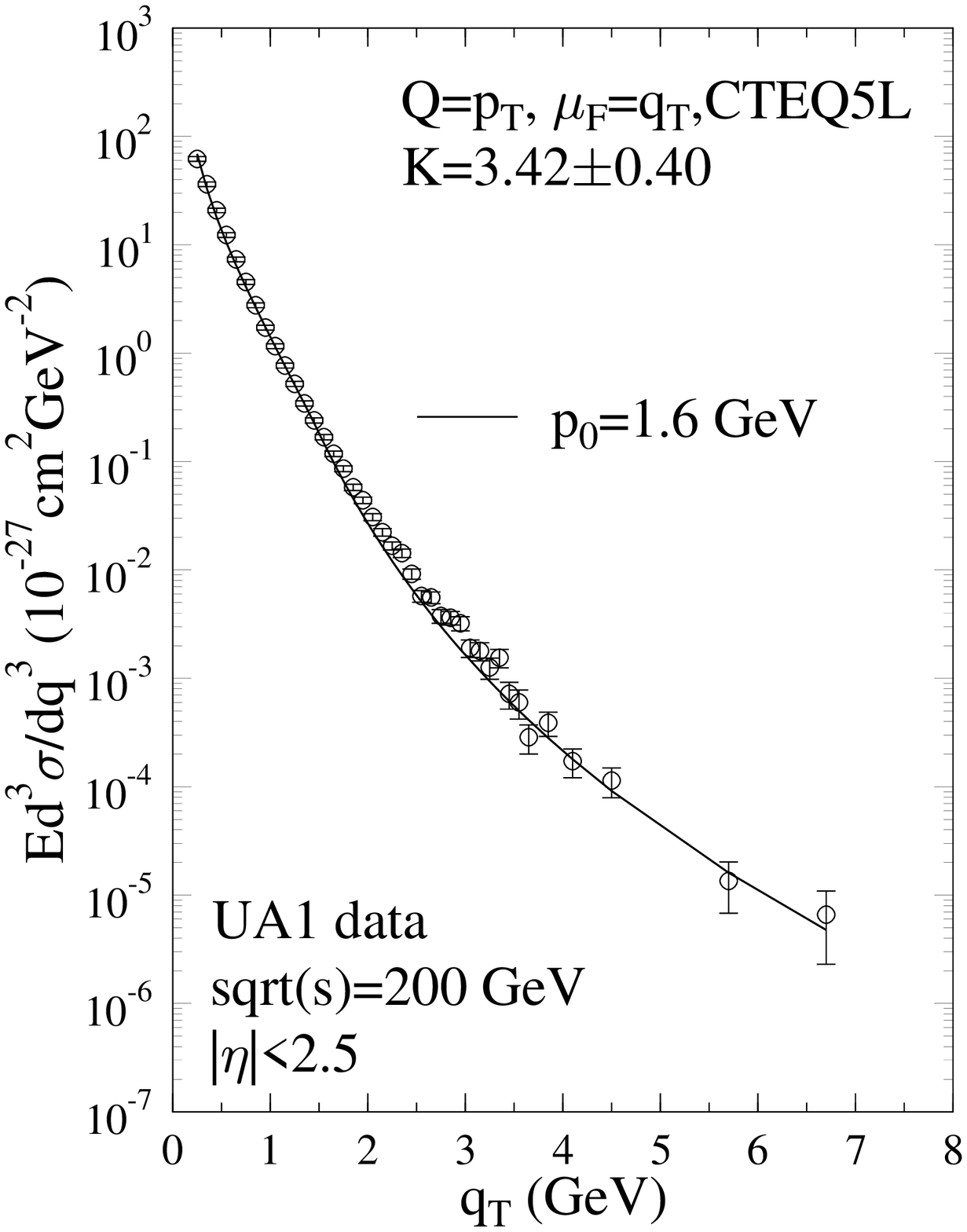} \hspace{-1.0cm}
\epsfysize=13.5cm\epsffile{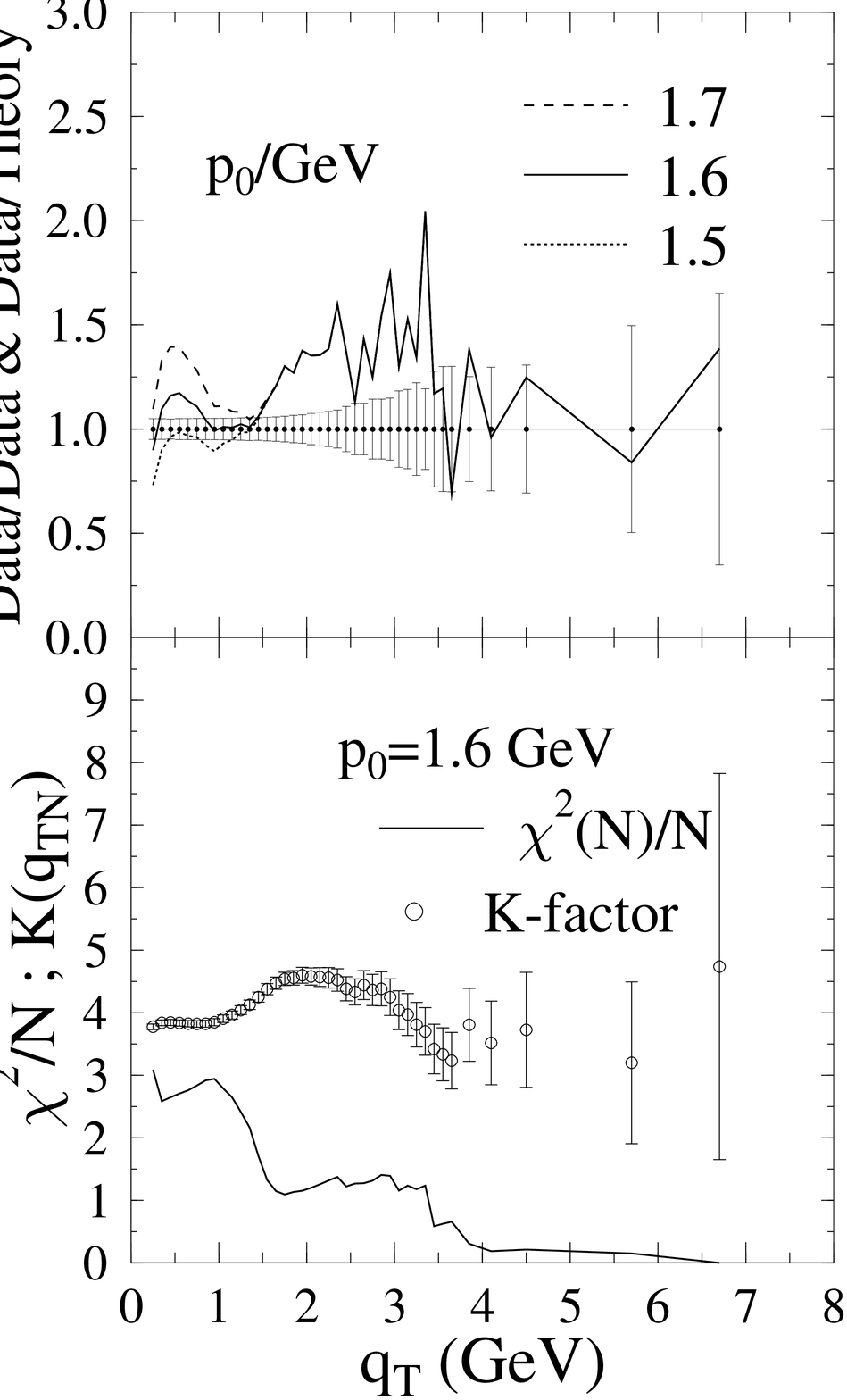} }
\vspace{-2cm}
\caption[a]{\protect \small As Fig.~\ref{mitt900} but for
$\sqrt{s}=200$ GeV. 
}
\label{mitt200}
\end{figure}

In Figs.~\ref{mitt900}, \ref{mitt500} and \ref{mitt200} we present the
computed $q_T$-spectra which correspond to the UA1 data \cite{UA1} at
$\sqrt{s}=900$, 500 and 200 GeV for $p +\bar p\to h + X$, where
$h\equiv (h^++ h^-)/2.$ The data and the computed spectra are
averaged over $\vert\eta\vert <2.5$. The systematic error for the
data is $\pm 15\%$ (not shown). The overall agreement with the data for 
$\sqrt{s}=900$~GeV is poor for $q_T\leq 5$~GeV. 
A reasonable estimate of $p_0$ in this case is to take the value which at
least reproduces the data at smallest values of $q_T$. The result for
$p_0=1.6$~GeV with the data is shown in Fig.~\ref{mitt900} (left panel).
The results for $\sqrt{s}=500$ GeV are presented in Fig.~\ref{mitt500}. 
The agreement with the data is now reasonable. Also for $\sqrt{s}=200$ GeV, 
shown in Fig.~\ref{mitt200}, the agreement with the data is again tolerable.

\begin{figure}[f]
\vspace{-2.0cm} \centerline{\hspace{-2.cm}
\epsfysize=17cm\epsffile{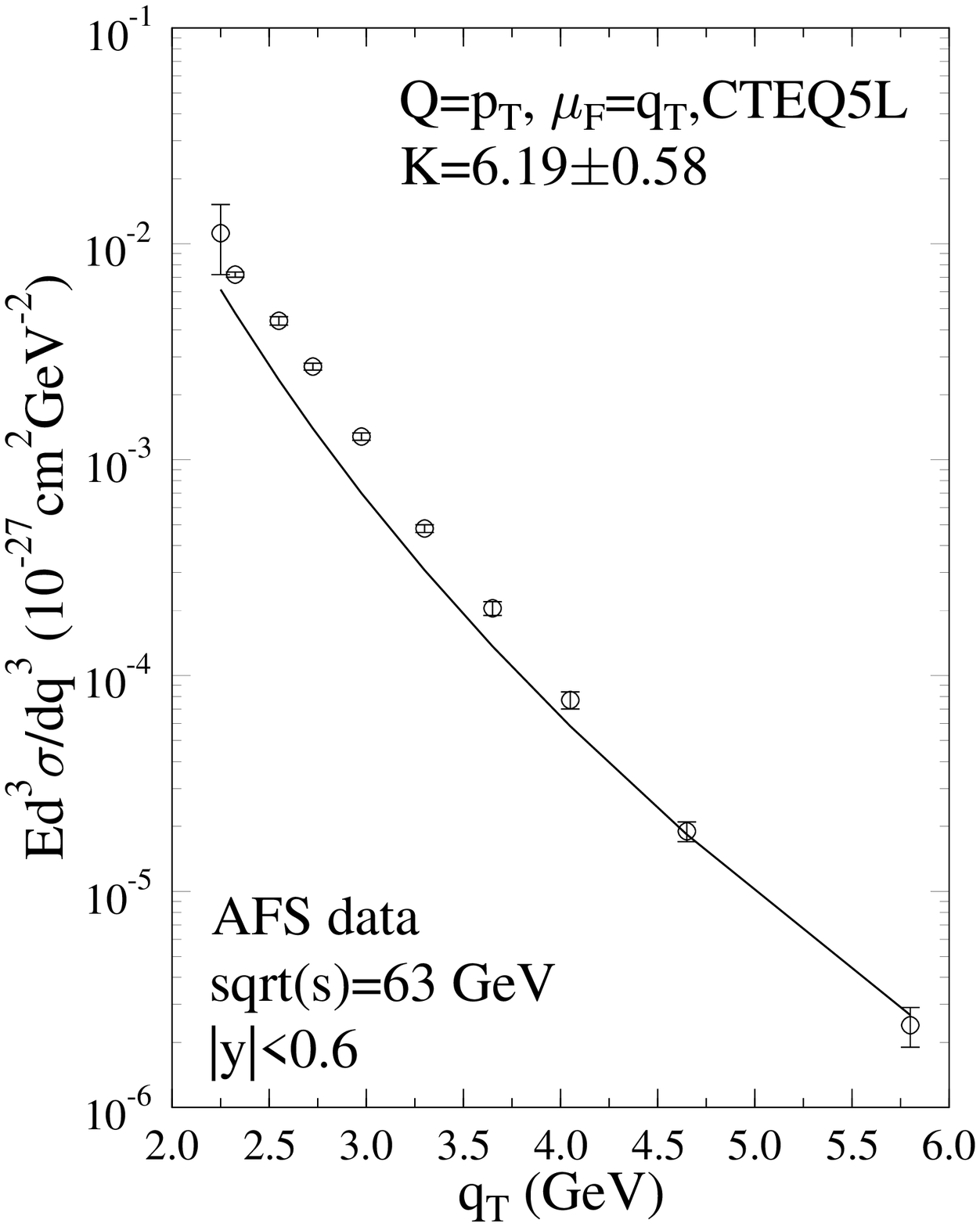} \hspace{-1.0cm}
\epsfysize=13.5cm\epsffile{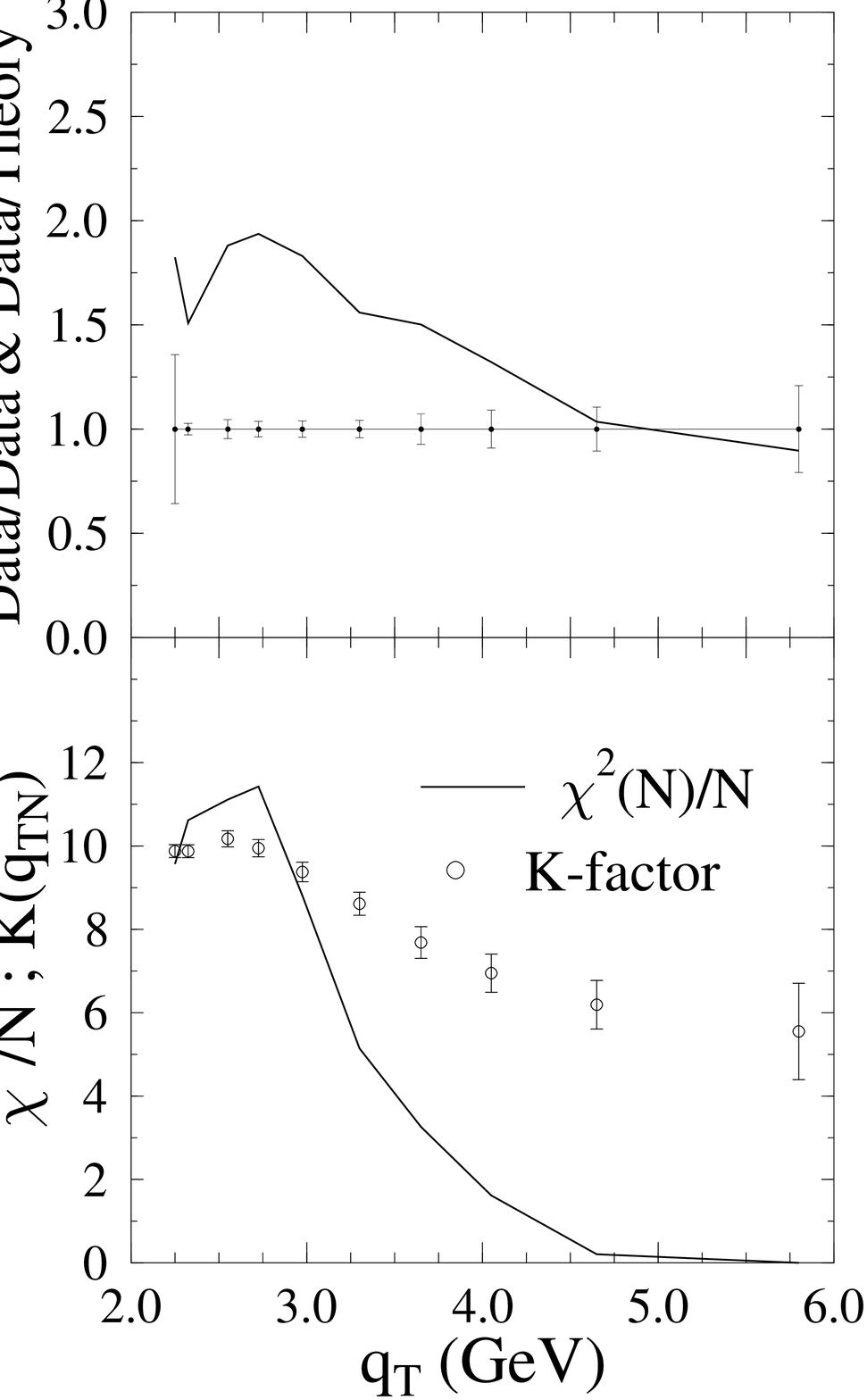} }
\vspace{-2cm}
\caption[a]{\protect \small As Fig.~\ref{mitt630_1} but for 
$h\equiv h^+ +h^-$, $\sqrt{s}=63$ GeV and $\vert y\vert <0.6$.
The data are from \cite{AFS}.
}
\label{mitt63}
\end{figure}

The last (and the smallest $\sqrt s$) dataset studied is the AFS 
data \cite{AFS} for  $p + p\to h + X$ ($h\equiv h^+ + h^-$) at 
$\sqrt{s}=63$ GeV, 
averaged over $\vert y\vert <0.6$. The systematic error for the data is 
$\pm 20\%$ (not shown). The results are illustrated  in Fig.~\ref{mitt63}. 
The data are for $q_T\geq 2.25$ GeV so they cannot be used to determine the 
value of $p_0$.

\begin{figure}[f]
\vspace{-2.0cm}
\centerline{\hspace{-2.cm} \epsfysize=17cm\epsffile{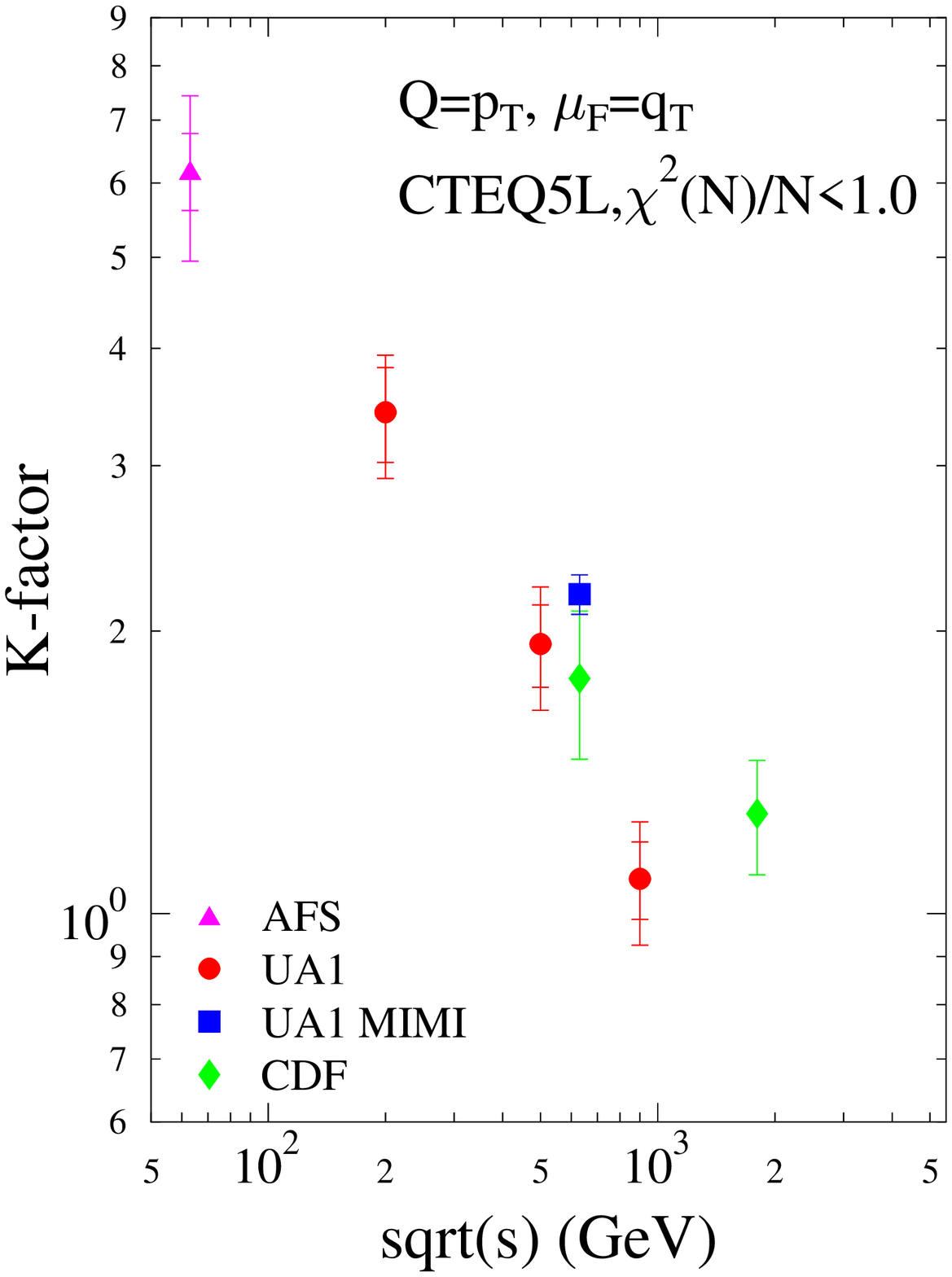} }
\vspace{-2cm}
\caption[a]{\protect \small The obtained $K$-factors for
$\sqrt{s}=63,200,500,630,900, 1800$ GeV. The inner error bars are
$\Delta K$ calculated from Eq.~(\ref{k-error}) and the outer error bars
(when shown) are the reported systematic errors for each data set
(see Table~1).  }
\label{koot}
\end{figure}

The $K$-factors obtained for different energies are gathered together in
Table~1 and in Fig.~\ref{koot}. The inner
error bars correspond to the statistical errors $\Delta K$ computed
from Eq.~(\ref{k-error}), and the outer error bars stand for the
systematic errors alone (when available). Except for the UA1 dataset
at $\sqrt s=900$ GeV, the obtained $K$-factors decrease systematically
with increasing $\sqrt s$. We also observe that a phenomenological fit
of the form $\log K=A+B\log(\sqrt s)$ can be justified. As 
the systematic error bars are not available for all cases, we do not perform such a 
fit here
but note that based on the numbers explicitly given in Table~1, such a
fit can be easily performed.

\section{AA-collisions; nuclear effects}

Next we study inclusive charged-hadron production in nuclear
collisions.  Our emphasis is at the highest cms-energies and at the
highest transverse momenta. We do not make an attempt to invoke any
model for the Cronin effect which is observed in $pA$ and $AA$
collisions at lower energies \cite{CRONIN}. Different phenomenological
approaches to include the Cronin effect to the pQCD-motivated framework can be
found e.g. in \cite{XNW,LEVAI}. We do not include any modifications of
the fragmentation functions, either, which are expected to arise in a
very dense medium \cite{QUENCHING,Dmod,SW}.  Instead, aiming at pQCD
cross sections, where additional model details are kept in a minimum,
we remain within the collinearly factorized leading-twist framework
where the nuclear effects in hadron production arise only from the nuclear
modifications of the parton distributions functions.

We define the average number density distribution of a parton flavour
 $i$ in a nucleus $A$ as
\begin{equation}
Af_{i/A}(x,Q^2)\equiv  	Zf_{i}^{p/A}(x,Q^2) + (A-Z)f_{i}^{n/A}(x,Q^2),
\label{nPDF1}
\end{equation}
where the nuclear parton distributions (nPDF) are defined 
in terms of the corresponding distributions $f_i$ in the free proton as 
\begin{equation}
f_{i}^{p/A}(x,Q^2)\equiv R_i^A(x,Q^2)f_i(x,Q^2).
\label{nPDF2}
\end{equation}
The nuclear modifications $R_i^A$ depend on the nucleus, the parton flavour, 
$x$, and through the DGLAP evolution \cite{DGLAP} of $f_{i}^{p/A}$ also on the
scale $Q^2$.  For $R_i^A(x,Q^2)$ we use the EKS98-parametrization
\cite{EKS98} which is based on a global DGLAP analysis of the nPDF
\cite{EKR98}. The distributions in bound neutrons are obtained through
isospin symmetry; $f_{u(\bar u)}^{n/A}=f_{d(\bar d)}^{p/A}$ and
$f_{d(\bar d)}^{n/A}=f_{u(\bar u)}^{p/A}$. We do not consider the
impact parameter dependence of the nPDF \cite{KJE,RAMONA} here.  Clearly,
there are two kinds of nuclear effects in the inclusive nuclear cross
sections to be computed: effects due to the nuclear modifications of
parton distributions (let us call this shadowing for brevity) and
isospin effects.

\begin{figure}[f]
\vspace{-1.5cm}
\centerline{\hspace{-1.cm} \epsfysize=10cm\epsffile{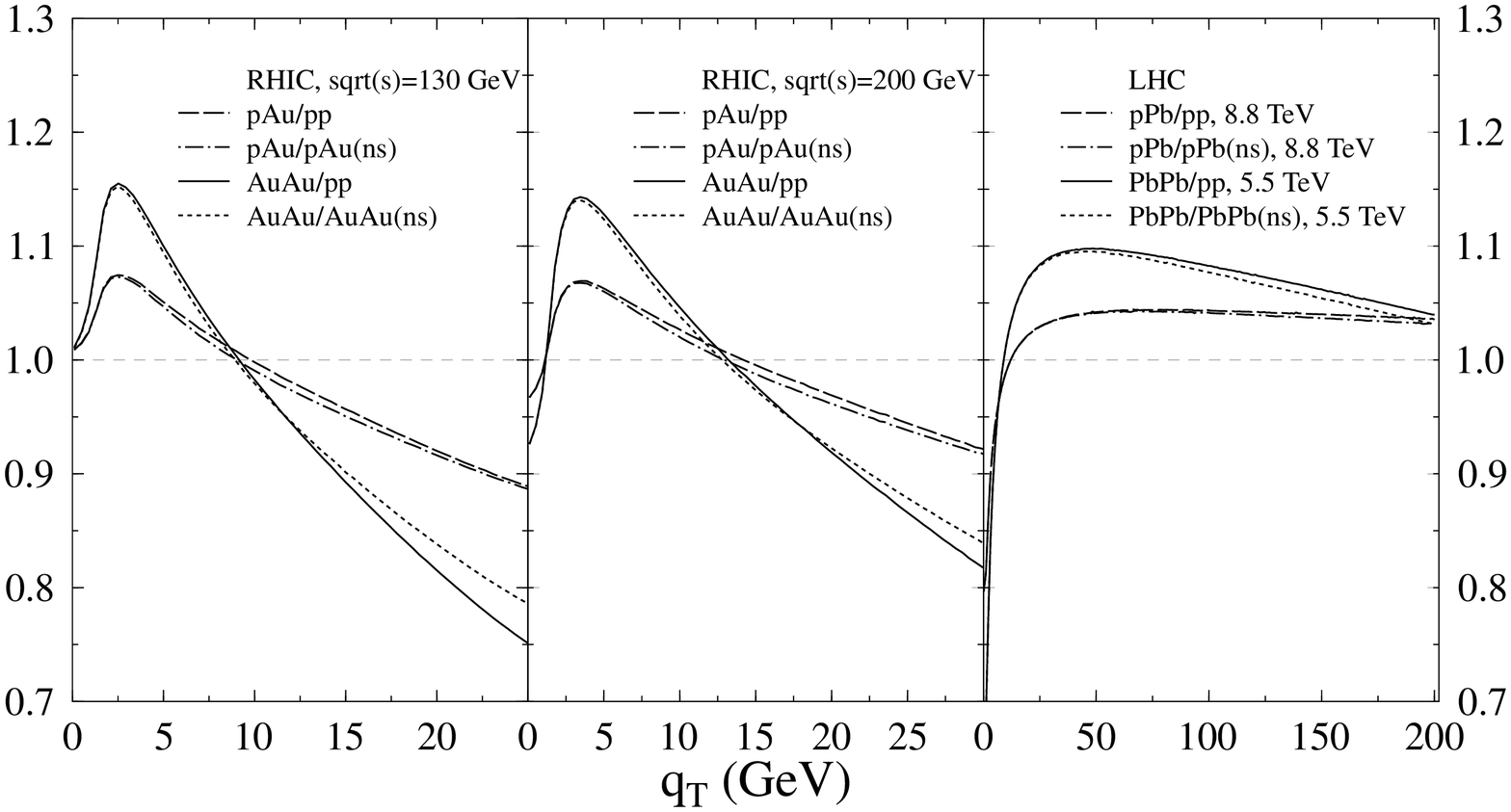} }
\vspace{-1.5cm}
\caption[a]{\protect \small The ratios of transverse momentum spectra
of charged hadrons in $AA(pA)$ collisions relative to $pp$ collisions
and $AA(pA)$ collisions without shadowing. At RHIC both Au+Au and $p$+Au
are at $\sqrt{s}=130$~GeV and 200 GeV.  At the LHC Pb+Pb is at 
$\sqrt{s}=5.5$~TeV and $p$+Pb at $\sqrt{s}=8.8$ TeV. }
\label{shadowing}
\end{figure}

The inclusive cross sections at $y=0$ in $pA$ and $AA$ collisions at RHIC and
LHC energies are computed from Eq.~(\ref{dsigma/dqT2}) with the nPDF
defined above.  Dividing the obtained cross sections by the ones
which do not include shadowing but include isospin effects, we obtain
the ratios shown in Fig.~\ref{shadowing} by the dotted and dotted-dashed
curves.  In order to see the magnitude of the isospin effects, we
compare the nuclear cross sections computed with shadowing and isospin
effects against the ones computed for $pp$. The resulting ratios,
$R_{pA}(q_T)$ and $R_{AA}(q_T)$ are shown by the dashed and solid lines. 
We can see that the isospin effects remain quite small in all
cases. 

The excess in the ratios $R_{pA}(q_T)$ and $R_{AA}(q_T)$ in
Fig.~\ref{shadowing} is caused by antishadowing in the nPDF (see
Fig.~3 of Ref. \cite{EKS98}). The depletion at small values of $q_T$
is due to shadowing at small $x$, and the depletion at larger values
of $q_T$ is due to the EMC effect in the nPDF. The systematics in the
location of the excess in $q_T$, its magnitude (the height is
decreasing with $\sqrt s$) is easy to understand by the following
argument.

\begin{figure}[f]
\vspace{-1.5cm}
\centerline{\hspace{-1.cm} \epsfysize=15cm\epsffile{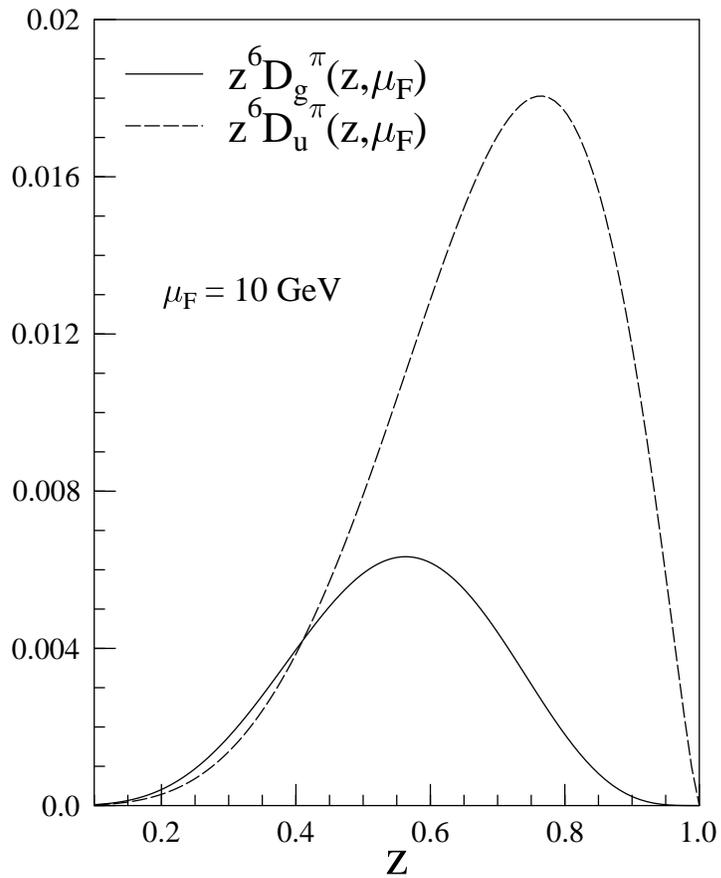} }
\vspace{-1.5cm}
\caption[a]{\protect \small 
The integrand of Eq.~(\ref{approks}) with $n=7$. The fragmentation 
functions are from KKP \cite{KKP}. The integrand is 
peaked at $\bar z_f$.
  }
\label{moments}
\end{figure}

Approximating the partonic cross section roughly by a power-law, 
$d\sigma^f/dp_Tdy\approx C_fp_T^{-n}$, and neglecting all masses, 
Eq.~(\ref{dsigma/dqT2}) simplifies into
\begin{equation}
\frac{d\sigma}{dq_Tdy}^{\hspace{-0.3cm}NN\rightarrow h+X}\approx 
\sum_{f=g,q,\bar q} \frac{C_f}{q_T^n}
\int dz z^{n-1}D_{f\rightarrow h}(z,q_T^2),
\label{approks}
\end{equation}
For $\sqrt s=200$ GeV at $p_T\sim 10$~GeV, we find $n\sim7$ and
somewhat less for smaller values of $p_T$ or larger values of $\sqrt
s$.  From Fig.~\ref{moments} for $z^6D(z,q_T^2)$ at $q_T=10$ GeV, we
observe that the integrand is strongly peaked around $\bar
z_f=0.6\dots0.8$, depending on the flavour of the fragmenting
parton. Obviously then $\int dz z^{n-1}D_{f\rightarrow h}(z)\approx
\bar z_f^{n-1}D_{f\rightarrow h}(\bar z_f,q_T^2)$. Consider only gluons for
simplicity (gluons dominate parton production when $x\ll1$ but this is
to some extent compensated by the larger fragmentation functions of
the quarks).  We find that $d\sigma^h/dq_Tdy\sim d\sigma^g/dp_Tdy$
where $p_T\approx q_T/\bar z$ with $\bar z\approx0.6$.

At $y=0$, the inclusive parton production dominantly probes the
region $x\sim 2p_T/\sqrt s$.  The ratio of shadowed and non-shadowed
cross sections in Fig.~\ref{shadowing} thus behaves as
\begin{equation}
R_{AA}(q_T,\sqrt s)
\sim [R_g^A(\frac{2q_T}{\bar z\sqrt s}, \frac{q_T}{\bar z})]^p, 
\label{RgA}
\end{equation}
where, due to the smearing caused by the integration over $y_2$,  
an effective power 1$<$$p$$<$2 appears.
Comparison with Fig.~3 of Ref. \cite{EKR98} for $R_g^A(x,Q^2)$, shows
that this is indeed the case: the cross-over points in $R_{AA}(q_T)$ can be
traced back to the cross-over points of the ratio $R_g^A$ in Eq.~(\ref{RgA}).
The decrease in the magnitude of the excess in $R_{AA}$ with growing
$\sqrt s$ is due to the scale evolution of $R_g^A$: at higher scales
the antishadowing of $R_g^A$ decreases.  Notice also the difference in
$\sqrt s$ for $pA$ and $AA$ at the LHC.

\section{Comparison with the PHENIX data}

As our final task - armed with the $\sqrt s$ dependence of the
$K$-factors - we compare our results with the transverse momentum
distributions of charged hadrons in Au+Au collisions at $\sqrt s=130$
GeV at RHIC, which have been recently measured by PHENIX
\cite{PHENIX}. We consider an average quantity
\begin{equation}
\bigg\langle\frac{dN^{AA\rightarrow h+X}}{d^2q_Td\eta}\bigg\rangle_c
=
\bigg\langle T_{AA}(b)
\frac{d\sigma^{NN\rightarrow h+X}}{d^2q_Td\eta}\bigg\rangle_c
=\langle N_{\rm binary}^{AA}\rangle_c \frac{1}{\sigma_{\rm
in}^{NN}}\frac{d\sigma}{d^2q_Td\eta}^{\hspace{-0.2cm}NN\rightarrow h+X},
\label{dNdpT}
\end{equation}
where $T_{AA}$ is the standard nuclear overlap function, $NN$ refers
to nucleon-nucleon collisions, and $\langle N_{\rm
binary}^{AA}\rangle_c$ is the average number of binary collisions
within the centrality selection $c$.  In Ref.~\cite{PHENIX}, two
centrality classes are given, $c_1=0\dots10$\% and $c_2=60\dots80$\%,
along with the estimates $\langle N_{\rm binary}\rangle_{c_1}=905\pm
96$ and $\langle N_{\rm binary}\rangle_{c_2}=20\pm6$ (resulting from a
Glauber analysis) and for the inelastic nucleon-nucleon cross section
$\sigma_{\rm in}^{NN}=40\pm3$ mb. We use these values here. To roughly
estimate the impact-parameter dependence of shadowing, we find an
effective nucleus $A_{\rm eff}$ for which the number of binary
collisions in a central $A_{\rm eff}A_{\rm eff}$ collision equals
$\langle N_{\rm binary}\rangle_{c_2}$. With $T_{AA}(0)\approx A^2/\pi
R_A^2$, this leads to $A_{\rm eff}^{c_1} \approx 163$ and $A_{\rm
eff}^{c_2}\approx 9$ for the two centrality classes studied.
Shadowing is included according to Eqs.~(\ref{nPDF1}) and (\ref{nPDF2}) 
for $A=A_{\rm eff}$. Thus the nuclear effects for the sample $c_1(c_2)$ 
remain within 15(5)\% at $q_T\leq 10$ GeV.

\begin{figure}[f]
\vspace{-1.5cm}
\centerline{\hspace{-1.cm} \epsfysize=15cm\epsffile{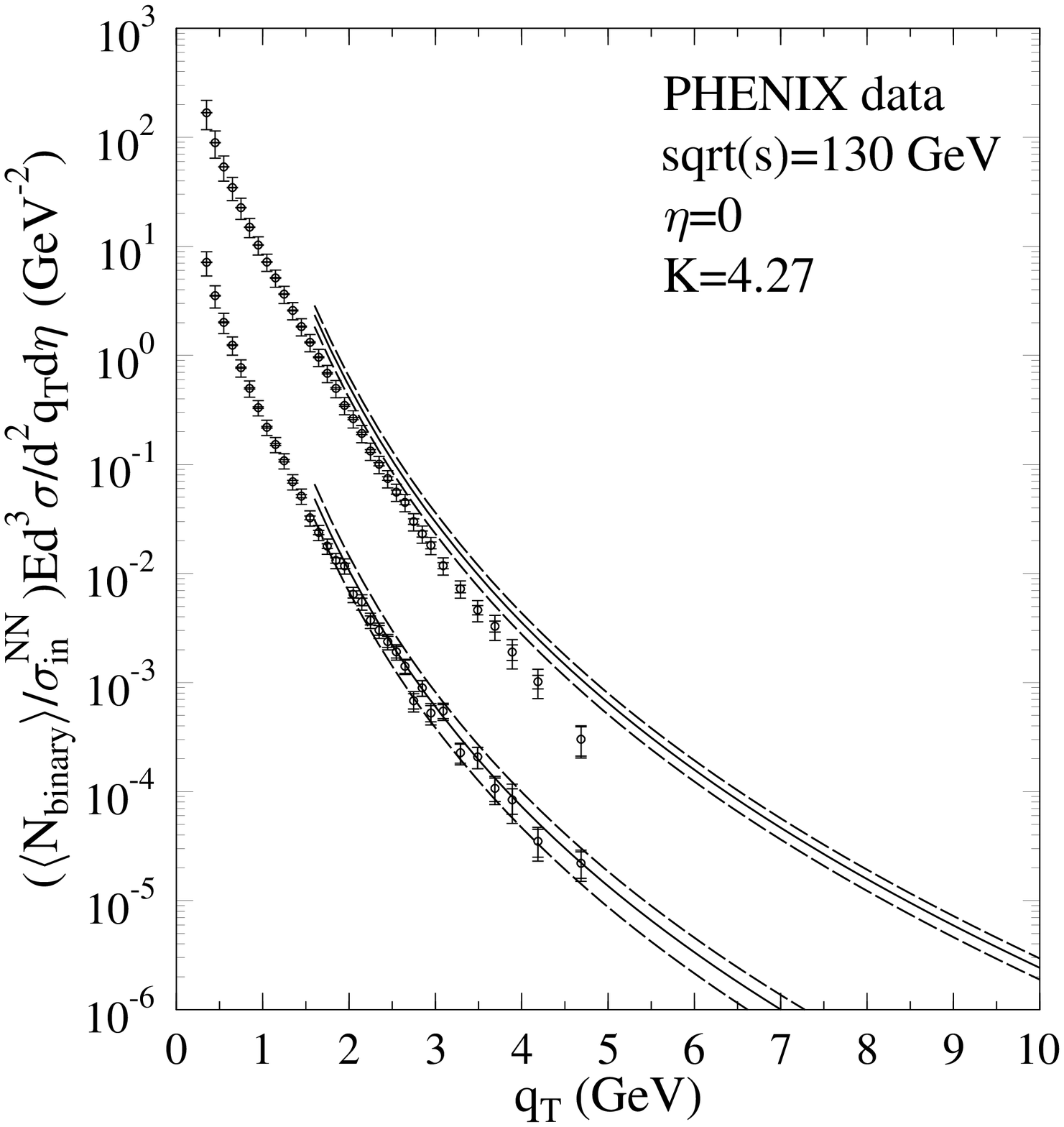} }
\vspace{-1.5cm}
\caption[a]{\protect \small Charged-hadron spectra for
centrality classes 0...10\% and 60...80\% in Au+Au collisions at
$\sqrt s=$ 130 $A$GeV at RHIC. The data are from PHENIX
\cite{PHENIX}. The solid curves are the pQCD results computed with
$K=4.27$ and $A_{\rm eff}=163$ and 9 for the two centrality
selections. The error bands shown by the dashed lines correspond to
an estimated total uncertainty of $\pm$22\% and $\pm$36\%, respectively 
(see the text).  }
\label{phenix}
\end{figure}

From the systematics of the $\sqrt s$-dependence of the $K$-factors in
Fig.~\ref{koot} it is observed that a simple power-law interpolation
between $\sqrt s=63$ GeV and 200 GeV gives a fair first estimate of
the $K$-factor at 130 GeV. This results in $K(\sqrt s=130\,{\rm GeV})=
4.27\pm0.18$.  For the error estimate, we have added the statistical
and systematic errors of $K(\sqrt s=200\, {\rm GeV})$ in quadrature.
Applying this $K$-factor, we plot the hadron spectra shown by the
solid curves in Fig.~\ref{phenix}.  For obtaining an estimate of the
total relative error (shown in the figure with the dashed lines), we 
add the relative errors of $K$, $\langle N_{\rm
binary}^{AA}\rangle_c$, and $\sigma_{\rm in}^{NN}$ in quadrature.
This leads to $\pm$22\% and $\pm$36\% for the central and peripheral 
cases.

A very good agreement is found between the computed cross sections and
the peripheral ($c_2$) data at the highest $q_T$ measured.  The
computed spectrum remains also surprisingly close to the measured one
at $q_T\sim 2$ GeV which was the trouble region in the $p\bar p(p)$ case. The
estimated error band now covers the most of expected magnitude of the
deviations (see Fig.~\ref{mitt200}).  Within the uncertainties 
 estimated, the
peripheral Au+Au collisions thus seem to behave effectively as $NN$
collisions.  In central collisions, however, especially towards the
largest transverse momenta, the computed spectrum systematically lies
above the data. We emphasize that it is in this region that our
approach should work the best. It should also be noted that if the
Cronin effect for high-$p_T$ parton production becomes more important
at central Au+Au collisions than in the peripheral case, it should
{\em increase} the computed cross sections.  This would in
turn make the difference between the measured and computed spectra
even larger from what is seen in Fig.~\ref{phenix}.  The high-$q_T$
hadron production in central Au+Au collisions thus looks quite
different from a straightforward extrapolation of the $pp$ case.

\section{Discussion}
In this paper we have studied inclusive charged-hadron production in
$p\bar p(p)$ collisions at $\sqrt s\ge 63$ GeV in the framework of
leading-twist lowest-order perturbative QCD. Modern lowest-order sets
of parton distributions and fragmentation functions are
used. Comparison with the AFS \cite{AFS}, UA1 \cite{UA1}, UA1 MIMI
\cite{MIMI}, and CDF \cite{CDF} data is made, and the difference
between the computed and measured spectra is quantified in terms of a
factor $K=d\sigma^{\rm exp}/d\sigma_{\rm LO}^{\rm th}$.  We have extracted
the $K$-factors and their $\sqrt s$-dependence from the highest
transverse momenta of the measured spectra.  A systematic decrease of
$K$ with growing $\sqrt s$ is found.  Also error estimates for the
obtained $K$ are given. We emphasize that the values of $K$ depend on
the PDFs, fragmentation functions and the scale choices used. 
The results of our analysis are useful for
further studies of hadron spectra, at cms-energies where no $pp$ data
are so far available.

We have also studied the lower limits for partonic transverse momentum
exchange, the cut-off scale $p_0$. A systematic increase of $p_0$ with
growing $\sqrt s$ is found and $p_0=1.6\dots2.2$ GeV for $\sqrt
s=63\dots1800$ GeV. It will be interesting to study the relation of
the scales $p_0$ obtained here to the saturation scales in $AA$
collisions \cite{EKRT}. For the models with semihard (perturbative) and
soft (nonperturbative) components for particle production, our analysis thus
suggests a cut-off scale $p_0$ which increases with $\sqrt s$.  The
same observation was also made e.g. in a recent analysis for the
HIJING model \cite{newHIJING}. Notice that the values of $p_0$ we
obtain are somewhat below those obtained in \cite{newHIJING}, since here 
we have described all particle production with the perturbative component 
only.

Once the $K$-factor in each case has been determined, we find the
overall agreement between the computed and measured spectra in $p\bar
p$ at $\sqrt s\ge200$ GeV relatively good (except for the UA1 dataset
at 900 GeV).  In general, the fragmentation functions of gluons
are not as well constrained by the $e^+e^-$ data as those of quarks and
antiquarks. In the future it will be very interesting to see whether
the agreement of the collinearly factorized cross sections and the
data will be improving with new sets of fragmentation functions.
It will also be interesting to see how close the results obtained here are 
with the scale-optimized NLO results \cite{AURENCHE} at various energies.

The agreement with the $p\bar p$ data at high transverse momenta gives
us confidence in that a reference cross section can be obtained by
making the extrapolation from $pp$ to $AA$ collisions.  The computed
spectra are compared with the recent measurements by PHENIX
\cite{PHENIX}.  We have shown that in the region $q_T=1...10$ GeV for
central Au+Au collisions at RHIC the antishadowing effects in parton
distributions {\em enhance} the spectra but only by less than 15\%. 
We have also shown that the isospin effects are small.  Using the
estimated $K=4.27$ for $\sqrt s=130$ GeV, together with the results
from a Glauber analysis of PHENIX, we have computed the inclusive
hadron spectra corresponding to the peripheral and central data
samples.  The results found are very similar to those obtained by
PHENIX: the peripheral collisions seem to behave practically as $pp$
collisions, and the computed spectrum agrees nicely with the data at
the largest values of $q_T$.  In comparison with the data with a
0..10\% centrality cut, the computed spectrum is found to
systematically lie above the measured spectrum at the highest
transverse momenta, where the emphasis of our approach is. An
additional Cronin effect, if important at all at this high $\sqrt s$
and large $q_T$, can be expected to enhance the computed spectrum and
thus enhance the deviation from the data. In order to see the effect
even more clearly, we are looking forward to more data points at even
higher values of $q_T$.

The very dense partonic medium produced (see the initial conditions e.g. 
in \cite{EKRT}) can be expected to be responsible for the
suppression of the high-$q_T$ hadrons \cite{QUENCHING}. There is an
increasing activity to study the fragmentation functions modified by
the presence of the partonic medium \cite{Dmod,SW,BDMS}. Incorporating
such modifications into the present analysis will be done next
\cite{EHSW}. This procedure focuses on the large-$q_T$ region,
where perturbative methods apply.  The other extreme case, emphasizing
the smaller-$q_T$ region, is hadron production from a fully thermalized
system described in terms of relativistic hydrodynamics \cite{ERRT}. 
It will be very interesting to study how these two regions merge to form the 
measured spectrum.

\subsection*{Acknowledgements}
We thank V. Karim\"aki, V. Ruuskanen, C. Salgado and U. Wiedemann for
discussions.  Financial support from the Academy of Finland, grants
no. 50338 and 163065, is gratefully acknowledged.

\end{document}